\documentclass[aps,pr,twocolumn,showpacs,superscriptaddress,groupedaddress,nofootinbib]{revtex4}  

\usepackage{longtable}
\usepackage{graphicx}  
\usepackage{ulem}   
\usepackage{comment} 
\usepackage{datetime}
\usepackage{dcolumn}

\usepackage{amsxtra}
\usepackage{euscript} 
\usepackage{bm}        
\usepackage{tabularx}
\usepackage{float}
\restylefloat{table}

\usepackage[cmtip,arrow]{xy}
\usepackage{pb-diagram,pb-xy}

\usepackage{amsmath}
\usepackage{amssymb}
\usepackage{amsthm}
\usepackage[pdftex]{color}
\usepackage[pdftex,colorlinks,citecolor=blue,linkcolor=blue,urlcolor=blue]{hyperref} 
\usepackage{graphicx}
\usepackage{dcolumn} 
\usepackage{bm} 

\usepackage{longtable}

\usepackage{ulem}   
\usepackage{comment} 
\usepackage{datetime}

\usepackage{tabularx}
\usepackage{float}
\restylefloat{table}

\newcommand{\beq}{\begin{equation}}
\newcommand{\eeq}{\end{equation}}
\newcommand{\bea}{\begin{eqnarray}}
\newcommand{\eea}{\end{eqnarray}}
\newcommand{\barr}{\begin{array}}
\newcommand{\earr}{\end{array}}

\long\def\begincomment#1\endcomment{}

\newcommand{\SU}{\mathrm{SU}}
\newcommand{\SO}{\mathrm{SO}}
\newcommand{\U}{\mathrm{U}}
\newtheorem{theorem}{Theorem}
\newtheorem{definition}{Definition}

\newtheorem{proposition}{Proposition}

\newtheorem{claim}{Claim}

\newcommand{\perm}{\mathrm{perm}}
\newtheorem{lemma}{Lemma}

\usepackage{pgf,tikz}
\usetikzlibrary{arrows}

\usepackage{bm}
\usepackage{bbold}

\usepackage{mathtools}

\DeclarePairedDelimiterX\braket[2]{\langle}{\rangle}{#1 \delimsize\vert #2}

\begin{document}


\title{Pedagogical comments about nonperturbative Ward-constrained melonic renormalization group flow
}

\author{Vincent Lahoche} \email{vincent.lahoche@cea.fr}   
\affiliation{Commissariat à l'\'Energie Atomique (CEA, LIST),
 8 Avenue de la Vauve, 91120 Palaiseau, France}

\author{Dine Ousmane Samary}
\email{dine.ousmanesamary@cipma.uac.bj}
\affiliation{International Chair in Mathematical Physics and Applications (ICMPA-UNESCO Chair), University of Abomey-Calavi,
072B.P.50, Cotonou, Republic of Benin}
\affiliation{Commissariat à l'\'Energie Atomique (CEA, LIST),
 8 Avenue de la Vauve, 91120 Palaiseau, France}

\date{\today}

\begin{abstract}
This paper, in addition to our recent works, intends to explore the behavior of the Wetterich flow equations in the portion of the theory space spanned by non-branching melons constrained  with Ward-identities. We focus on a rank-5 just-renormalizable tensorial group field theory and consider a non-trivial extension of the local potential approximation namely effective vertex expansion for just-renormalizable quartic melonic interactions, disregarding effects coming from disconnected interactions. Investigating the dynamically constrained flow, we show explicitly that results weakly rely on the number of quartic interactions involved in the classical action. In particular, the predictions for the fully connected model are essentially the same as for the single colored model. Finally, closing the flow equations using Ward identities without additional assumptions to compute integrals involved in the effective vertex expansion, we do not find reliable fixed point in the unconstrained theory space  connected with the Gaussian region.
\end{abstract}


\maketitle

\section{Introduction} \label{sec1}
Recently, several investigations have been done concerning the study of the functional renormalization group (FRG) applied to tensor models (TM) and group field theory (GFT). The study of the FRG applied to tensorial group field theory (TGFT) is motivated by the expectation that phase transitions may occurs in the early Universe. However, TGFTs distinguish from ordinary quantum field theory by a specific notion of non-locality, requiring new computation tools to address the question of the renormalization group \cite{Lahoche:2019vzy}-\cite{Benedetti:2015yaa} in the  Wetterich-Morris framework \cite{Wetterich:1991be}-\cite{Wetterich:1992yh}. To deal with nonlocality of the theory, the first application focused on the standard vertex expansion with crude truncation in the theory space. Recently a new method called effective vertex expansion (EVE) \cite{Lahoche:2019vzy}-\cite{Lahoche:2018vun} has been introduced as a powerful tool to go beyond the vertex expansion and to keep the full momentum dependence of the effective vertex functions. Another important improvement concern the use of Ward identities (WI) \cite{Ward:1950xp}-\cite{Takahashi:1957xn}. The WI  is an additional constraint on the flow and therefore should not be overlooked in the study of the renormalization group. In particular a lot of allowed phase transitions which are identified near the fixed point are shown to be nonphysical due to the violation of the  \cite{Lahoche:2018ggd}. Note that the nontrivial form of the Ward identity for the TGFT with nontrivial propagator in the functional actions is not a consequence of the regulator $r_k$ but rather is due to the Laplacian in the kinetic action defining the theory. Let us remark that for standard gauge-invariant theories like QED see \cite{Gies:2006wv} the regulator breaks generally the explicit invariance of the kinetic term and leads to a new nontrivial Ward identity that depends on the regulator $r_k$. This is not the case for TGFT models for which the kinetic term intrinsically violates the global Unitary symmetry. Therefore the appearance of the regulator generalizes the definition of the theory but does not add any new information concerning the Ward identity.
\medskip

This present paper is a complement to our recent works on FRG applied to TGFT \cite{Lahoche:2019vzy}-\cite{Lahoche:2018vun}. It is organized as follows. In section \eqref{sec2} we provide in detail useful ingredients for the description of the FRG to TGFT. In the section \eqref{EVE} the EVE is derived to thank about the FRG with a new alternative way without truncation. The corresponding flow equations which improve the truncation method are given. Section \eqref{melons} describes our new proposal to merge the Wetterich equation and the Ward identity in the constrained melonic phase space $\mathcal{E}_{\mathcal C}$ of theory space. In section \eqref{COnst} we investigate the local potential approximation. Section \eqref{secfin} is devoted to another point of view in the computation of the scale dynamics called Callan Symanzik equation. In the last section \eqref{conclusion} we give our conclusion. 

\section{Preliminaries}\label{sec2}
A \textit{group field} $\varphi$ is a field, complex or real, defined over $d$--copies of a group manifold $\mathrm{G}$ rather than on space time:
\begin{equation}
\varphi: \mathrm{G}^d\to\mathbb{R},\mathbb{C}\,.
\end{equation}
Standard choices to make contact with physics are $\SU(2)$ and $\SO(4)$ \cite{Boulatov:1992vp}-\cite{Perez:2002vg}. In this paper, we focus only on the non-local aspects of the interactions, and consider the Abelian version of the theory, setting $\mathrm{G}=\U(1)$. For this choice, the field may be equivalently described on the Fourier dual group $\mathbb{Z}^d$ by a \textit{tensor field} $T:\mathbb{Z}^d\to \mathbb{C}$. We consider a theory for two complexes fields $\varphi$ and $\bar\varphi$, requiring two complex tensors fields $T$ and $\bar{T}$. The allowed configurations are then constrained by the choice of a specific action, completing the definition of the GFT. At the classical level, for free fields we choose the familiar form:
\begin{equation}
S_{\text{kin}}[T,\bar{T}]:=\sum_{\vec{p}\in\mathbb{Z}^d} \bar{T}_{p_1\cdots p_d}\left(\vec{p}\,^2+m^2\right) {T}_{p_1\cdots p_d}\,,\label{kin}
\end{equation}
with the standard notation $\vec{p}\,^2:=\sum_i p_i^2$, $\vec{p}:=(p_1,\cdots,p_d)$. For the rest of this paper we use the short notation $T_{\vec{p}}\equiv T_{p_1\cdots p_d}$. The equation \eqref{kin} defines the bare propagator $C^{-1}(\vec{p}\,):=\vec{p}\,^2+m^2$. Among the natural transformations that we can consider for a pair of complex tensor fields, the unitary transformations play an important role. They provide the principle that allows to build the interactions, which are chosen to be invariant under such a transformation. Denoting by $N$ the \textit{size} of the tensor field, restricting the domain of the indices $p_i$ into the window $[\![-N,N ]\!]$, we require invariance with respect to independent transformations along each of the $d$ indices of the tensors:
\begin{equation}
T^\prime_{p_1\cdots p_d}=\sum_{\vec{q}\in\mathbb{Z}^d} \left[\prod_{i=1}^d U^{(i)}_{p_iq_i}\right] \,T_{q_1\cdots q_d}\,,\label{unit}
\end{equation}
with $U^{(i)}(U^{(i)})^\dagger=\mathrm{id}$. Define $\mathbb{U}(N)$ as the set of unitary symmetries of size $N$, a transformation for tensors is then a set of $d$ independent elements of $\mathbb{U}(N)$, $\mathcal{U}:=(U_1,\cdots, U_d)\in\mathbb{U}(N)^d$, one per index of the tensor fields. The unitary symmetries admitting an inductive limit for arbitrary large $N$, we will implicitly consider the limit $N\to \infty$ in the rest of this paper \cite{Gurau:2011xp}.
\noindent
We call \textit{bubble} all the invariant interactions which cannot be factorized into two or more smaller bubbles. Observe that because the transformations are independent, the bubbles are not local in the usual sense over the group manifold $\mathrm{G}^d$. However, locality does not make sense without physical content. In standard field theory for instance, or in physics in general, the locality is defined by the way following which the fields or particles interact together, and as for tensors, this choice reflects invariance with respect to some transformations like translations and rotations. With this respect, the transformation rule \eqref{unit} define both the nature of the field (a tensor) and the corresponding locality principle. To summarize:
\begin{definition}
Any interaction bubble is said to be local. By extension, any functions expanding as a sum of bubble will be said local.
\end{definition}
This locality principle called \textit{traciality} in the literature has some good properties of the usual ones. In particular it allows to define local counter-terms and to follow the standard renormalization procedure for interacting quantum fields with UV divergences. In this paper, we focus on the quartic melonic model in rank $d=5$, describing by the classical interaction:
\begin{equation}
S_{\text{int}}[T,\bar T]= g \sum_{i=1}^d \vcenter{\hbox{\includegraphics[scale=0.8]{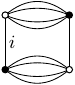} }}\,,\label{int}
\end{equation}
$g$ denoting the coupling constant and where we adopted the standard graphical convention \cite{Gurau:2011xp} to picture the interaction bubble as $d$-colored bipartite regular connected graphs. The black (resp. white) nodes corresponding to $T$ (resp. $\bar{T}$) fields, and the colored edges fixing the contractions of their indices. Note that, because we contract indices of the same color between $T$ and $\bar{T}$ fields, the unitary symmetry is ensured by construction. The model that we consider has been showed to be \textit{just renormalizable} is the usual sense, that is to say, all the UV divergences can be subtracted with a finite set of counter-terms, for mass, coupling and field strength. From now on, we will consider $m^2$ and $g$ as the bare couplings, sharing their counter-terms, and we introduce explicitly the wave function renormalization $Z$ replacing the propagator $C^{-1}$ by
\begin{equation}
C^{-1}(\vec{p}\,)=Z\vec{p}\,^2+m^2\,.
\end{equation}

\noindent
The equations \eqref{kin} and \eqref{int} define the classical model, without fluctuations. We quantize using path integral formulation, and define the partition function integrating over all configurations, weighted by $e^{-S}$:
\begin{equation}
\mathcal{Z}(J,\bar J):=\int dT d\bar T \,e^{-S[T,\bar T]+\langle \bar{J},T\rangle+\langle \bar{T},J\rangle}\,,\label{quantum}
\end{equation}
the sources being tensor fields themselves $J,\bar J:\mathbb{Z}^d\to \mathbb{C}$ and $\langle \bar{J},T\rangle:=\sum_{\vec{p}}\bar{J}_{\vec{p}}\, T_{\vec{p}}$. Note that the quantization procedure provide a canonical definition of what is UV and what is IR. The UV theory corresponding to the classical action $S=S_{\text{kin}}+S_{\text{int}}$ whereas the IR theory corresponds to the standard effective action defined as the Legendre transform of the free energy $\mathcal{W}:=\ln(\mathcal{Z}(J,\bar J))$. \\

\noindent
Renormalization in standard field theory allows to subtract divergences, and it has been showed that quantum GFT can be renormalized in the usual sense. Concerning the quantization procedure moreover, the renormalization group allows describing quantum effects \textit{scale by scale}, through more and more effective models, defining a path from UV to IR by integrating out the fluctuation of increasing size. \\

\noindent
Recognizing this path from UV to IR as an element of the quantization procedure itself, we substitute to the global quantum description \eqref{quantum} a set of models $\{\mathcal{Z}_k\}$ indexed by a referent scale $k$. This scale define what is UV, and integrated out and what is IR, and frozen out from the long distance physics. The set of scales may be discrete or continuous, and in this paper we choose a continuous description $k\in [0,\Lambda]$ for some fundamental UV cut-off $\Lambda$. There are several ways to build what we call functional renormalization group . We focus on the Wetterich-Morris approach \cite{Wetterich:1991be}-\cite{Wetterich:1992yh}, $ \mathcal{Z}_k(J,\bar J)$ being defined as:
\begin{equation}
\mathcal{Z}_k(J,\bar J):=\int dT d\bar T \,e^{-S_k[T,\bar T]+\langle \bar{J},T\rangle+\langle \bar{T},J\rangle}\,,\label{quantum2}
\end{equation}
with: $S_k[T,\bar T]:=S[T,\bar T]+\sum_{\vec{p}}\,\bar{T}_{\vec{p}}\, r_k(\vec{p}\,^2)T_{\vec{p}}$. The momentum dependent mass term $r_k(\vec{p}\,^2)$ called \textit{regulator} vanish for UV fluctuations $\vec{p}\,^2\gg k^2$ and becomes very large for the IR ones $\vec{p}\,^2\ll k^2$. Some additional properties for $r_k(\vec{p}\,^2)$ may be found in standard references \cite{Litim:2000ci}-\cite{Litim:2001dt}. Without explicit mentions, we focus on the Litim's modified regulator:
\begin{equation}
r_k(\vec{p}\,^2):=Z(k)(k^2-\vec{p}\,^2)\theta(k^2-\vec{p}\,^2)\,,\label{regulator}
\end{equation}
where $\theta$ designates the Heaviside step function and $Z(k)$ is the running wave function strength.   Note that this regulator imposes a bound on the value of the anomalous dimension at a given fixed point. Indeed, the regulator must behaves as $k^r$ with $r>0$ in the deep UV. With definition \eqref{regulator}: $r_k \approx Z k^2 = k^{2+\eta}$, requiring:
\begin{equation}
\eta > \eta_c:= -2\,. \label{physicalboundeta}
\end{equation}
\medskip

The renormalization group flow equation, describing the trajectory of the RG flow into the full theory space is the so called Wetterich equation \cite{Wetterich:1991be}-\cite{Wetterich:1992yh}, which for our model writes as:
\begin{equation}
\frac{\partial}{\partial k} \Gamma_k= \sum_{\vec{p}} \frac{\partial r_k}{\partial k}(\vec{p}\,) \left( \Gamma_k^{(2)}+r_k \right)^{-1}_{\vec{p}\,\vec{p}}\,,\label{Wett}
\end{equation}
where $(\Gamma_k^{(2)})_{\vec{p}\,\vec{p}\,^\prime}$ is the second derivative of the \textit{average effective action} $\Gamma_k$ with respect to the classical fields $M$ and $\bar{M}$:
\begin{equation}
\left(\Gamma_k^{(2)}\right)_{\vec{p}\,\vec{p}\,^\prime}=\frac{\partial^2\Gamma_k}{\partial M_{\vec{p}}\,\partial \bar{M}_{\vec{p}\,^\prime}}\,,
\end{equation}
where $M_{\vec{p}}=\partial \mathcal{W}_k/\partial\bar{J}_{\vec{p}}$, $\bar M_{\vec{p}}=\partial \mathcal{W}_k/\partial{J}_{\vec{p}}$ and:
\begin{align}
\nonumber\Gamma_k[M,\bar M]+\sum_{\vec{p}}\,\bar{M}_{\vec{p}}\, r_k(\vec{p}\,^2)M_{\vec{p}}&:=\langle \bar{M},J\rangle+\langle \bar J,M\rangle\\
&\,-\mathcal{W}_k(M,\bar M)\,,
\end{align}
with $\mathcal{W}_k=\ln(\mathcal{Z}_k)$.\\

\noindent
The flow equation \eqref{Wett} is a consequence of the variation of the propagator, indeed
\begin{equation}
\frac{\partial r_k}{\partial k} =\frac{\partial C^{-1}_k}{\partial k} \,,
\end{equation}
for the \textit{effective covariance} $C^{-1}_k:=C^{-1}+r_k$. But the propagator has other source of variability. In particular, it is not invariant with respect to the unitary symmetry of the classical interactions \eqref{int}. Focusing on an infinitesimal transformation : $\delta_1:=(\mathrm{id}+\epsilon,\mathrm{id},\cdots,\mathrm{id})$ acting non-trivially only on the color $1$ for some infinitesimal anti-Hermitian transformations $\epsilon$, the transformation rule for the propagator follows the Lie bracket:
\begin{equation}
\mathcal{L}_{\delta_1} C^{-1}_k= [C^{-1}_k,\epsilon]\,.
\end{equation}
The source terms are non invariant as well. However, due to the translation invariance of the Lebesgue measure $dT d\bar T$ involved in the path integral \eqref{quantum2}, we must have $\mathcal{L}_{\delta_1} \mathcal{Z}_k=0$. Translating this invariance at the first order in $\epsilon$ provide a non-trivial \textit{Ward-Takahashi identity} for the quantum model:
\begin{theorem} \textbf{(Ward identity.)}
The non-ivariance of the kinetic action with respect to unitary symmetry induce non-trivial relations between $\Gamma^{(n)}$ and $\Gamma^{(n+2)}$ for all $n$, summarized as:
\begin{align}
\nonumber\sum_{\vec{p}_\bot, \vec{p}_\bot^{\prime}}{\vphantom{\sum}}'
&\bigg\{\big[C_k^{-1}(\vec{p})-C_k^{-1}(\vec{p}\,^{\prime})\big]\left[\frac{\partial^2 \mathcal{W}_k}{\partial \bar{J}_{\vec{p}\,^\prime}\,\partial {J}_{\vec{p}}}+\bar{M}_{\vec{p}}M_{\vec{p}\,^\prime}\right]\\
&\quad-\bar{J} _{\vec{p}}\,M_{\vec{p}\,^\prime}+{J} _{\vec{p}\,^\prime}\bar{M}_{\vec{p}}\bigg\}=0\,.\label{Ward0}
\end{align}
where $\sum_{\vec{p}_\bot, \vec{p}_\bot^{\prime}}' :=\sum_{\vec{p}_\bot, \vec{p}_\bot^{\prime}} \delta_{\vec{p}\,\vec{p}_\bot^\prime}$.
\end{theorem}
In this statement we introduced the notations $\vec{p}_\bot:=(p_2,\cdots,p_d)\in\mathbb{Z}^{d-1}$ and $\delta_{\vec{p}\,\vec{p}_\bot^\prime}=\prod_{j\neq 1}\, \delta_{p_j\,p_j^\prime}$. Equations \eqref{Wett} and \eqref{Ward0} are two formal consequences of the path integral \eqref{quantum2}, coming both from the non-trivial variations of the propagator. Therefore, there are no reason to treat these two equations separately. This formal proximity is highlighted in their expanded forms, comparing equations \eqref{florence}--\eqref{florence2} and \eqref{W1}--\eqref{W2}. Instead of a set of partition function, the quantum model may be alternatively defined as an (infinite) set of effective vertices $\mathcal{Z}_k\sim \{\Gamma^{(n)}_k\}=:\mathfrak{h}_k$. RG equations dictate how to move from $\mathfrak{h}_k\underset{\text{RG}}{ \to}\mathfrak{h}_{k+\delta k}$ whereas Ward identities dictate how to move in the momentum space, along $\mathfrak{h}_k$.

\section{Effective vertex expansion}\label{EVE}
This section essentially summarizes the state of the art in \cite{Lahoche:2018hou}-\cite{Lahoche:2018vun}. The exact RG equation cannot be solved except for very special cases. The main difficulty is that the Wetterich equation \eqref{Wett} split as an infinite hierarchical system, the derivative of $\Gamma^{(n)}$ involving $\Gamma^{(n+2)}$, and so one. Appropriate approximation schemes are then required to extract information on the exact solutions. The effective vertex expansion (EVE) is a recent technique allowing to build an approximation considering infinite sectors rather than crude truncations on the full theory space. We focus on the \textit{melonic sector}, sharing all the divergences of the model and then dominating the flow in the UV. One recalls that melonic diagrams are defined as the diagram with an optimal degree of divergence. Fixing some fundamental cut-off $\Lambda$, we consider the domain $\Lambda \ll k \ll 1$, so far from the deep UV and the deep IR regime. At this time, the flow is dominated by the renormalized couplings, have positive or zero \textit{flow dimension} (see \cite{Lahoche:2018oeo}). We recall that the flow dimension reflects the behavior of the RG flow of the corresponding quantity, and discriminate between essential, marginal and inessential couplings just like standard dimension in quantum field theory\footnote{For ordinary quantum field theory, the dimension is fixed by the background itself. Without background, this is the behavior of the RG flow which fixes the canonical dimension.}. Because our theory is just-renormalizable, one has necessarily $[m^2]=2$ and $[g]=0$, denoting as $[X]$ the flow dimension of $X$. \\

Note that we focus on the strictly local potential approximation, in which the EVE method work well. As recalled in the previous section, locality for TGFT means that we can be expanded as a sum (eventually infinite) of interaction bubbles. In order to make contact the effective vertex formalism of this section, we have to recall the notion of boundary graphs. Indeed, effective vertex expand generally as a sum of connected diagrams, but what is relevant for locality is boundary locality, and the boundary map $\mathcal{B}$ is defined as follow:
\begin{definition}
Let $\mathfrak{G}_p$ the set of bubbles with at most $p$ black nodes. The boundary map $\mathcal{B} : \mathfrak{G}_p\to (\mathfrak{G}_p)^{\times p}$ is defined as follows. For any connected Feynman graph $\mathcal{G}$, $\mathcal{B}(\mathcal{G})$ is the subset of external nodes, linking together with colored edges, according to their connectivity path in the graph.
\end{definition}
\begin{figure}[h!]
\includegraphics[scale=0.9]{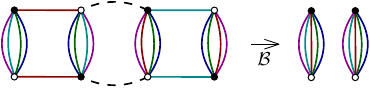} \label{figboundary}
\caption{Illustration of the manner that the boundary map work on a connected Feynman graph. The dotted edges correspond to Wick contractions. }
\end{figure}
We recall that external nodes are nodes hooked to external edges. Figure \ref{figboundary} provide an illustration of the definition. Any Feynman graph $\mathcal{G}$ contributing to a local effective vertex are then such that $\mathcal{B}(\mathcal{G})$ is a interaction bubble. \\
\noindent
The basic strategy of the EVE is to close the hierarchical system coming from \eqref{Wett} using the analytic properties of the effective vertex functions\footnote{Melonic diagrams may be easily counted as "trees", and the (renormalized) melonic perturbation series is easy to sum. } and the rigid structure of the melonic diagrams. More precisely, the EVE express all the melonic effective vertices $\Gamma^{(n)}$ having negative flow dimension (that is for $n>4$) in terms of effective vertices with positive or null flow dimension, that is $\Gamma^{(2)}$ and $\Gamma^{(4)}$, and their flow is entirely driven by just-renormalizable couplings. As recalled, in this way we keep the entirety of the melonic sector and the full momentum dependence of the effective vertices. \\
We work into the \textit{symmetric phase}, i.e. in the interior of the domain where the vacuum $M=\bar{M}=0$ make sense. This condition ensure that effective vertices with an odd number of external points, or not the same number of black and white external nodes have to be discarded from the analysis. These ones being called \textit{assorted functions}. Moreover, due to the momentum conservation along the boundaries of faces, $\Gamma^{(2)}_k$ must be diagonal:
\begin{equation}
\Gamma^{(2)}_{k,\,\vec{p}\,\vec{q}}=\Gamma_k^{(2)}(\vec{p}\,)\delta_{\vec{p}\,\vec{q}}\,.
\end{equation}
We denote as $G_k$ the effective $2$-point function $G^{-1}_k:=\Gamma_k^{(2)}+r_k$. \\
\noindent
The main assumption of the EVE approach is the existence of a finite analyticity domain for the leading order effective vertex functions, in which they may be identified with the resumed perturbative series. For the melonic vertex function, the existence of a such analytic domain is ensured, melons can be mapped as trees and easily summed. Moreover, these resumed functions satisfy the Ward-Takahashi identities, written without additional assumption than cancellation of odd and assorted effective vertices. One then expect that the symmetric phase entirely cover the perturbative domain. \\
\noindent
Among the properties of the melonic diagrams, we recall the following statement:
\begin{proposition}\label{propmelons}
Let $\mathcal{G}_N$ be a $2N$-point 1PI melonic diagrams build with more than one vertices for a purely quartic melonic model. We call external vertices the vertices hooked to at least one external edge of $\mathcal{G}_N$ has :
\begin{itemize}
\item Two external edges per external vertices, sharing $d-1$ external faces of length one.
\item $N$ external faces of the same color running through the interior of the diagram.
\end{itemize}
\end{proposition}
Due to this proposition, the melonic effective vertex $\Gamma^{(n)}_k$ decompose as $d$ functions $\Gamma^{(n),i}_k$, labeled with a color index $i$:
\begin{equation}
\Gamma^{(n)}_{k,\,\vec{p}_1,\cdots,\vec{p}_n}=\sum_{i=1}^d \Gamma^{(n),i}_{k,\,\vec{p}_1,\cdots,\vec{p}_n}\,.
\end{equation}
The Feynman diagrams contributing to the perturbative expansion of $\Gamma^{(n,i)}_{k,\,\vec{p}_1,\cdots,\vec{p}_n}$ fix the relations between the different indices. For $n=4$ for instance, we get, from proposition \ref{propmelons}:
\begin{equation}
\Gamma_{\vec{p}_1,\vec{p}_2,\vec{p}_3,\vec{p}_4}^{(4),i} = \vcenter{\hbox{\includegraphics[scale=0.5]{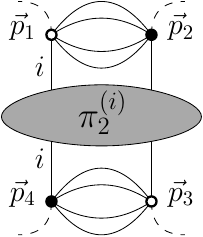} }}+\, \vcenter{\hbox{\includegraphics[scale=0.5]{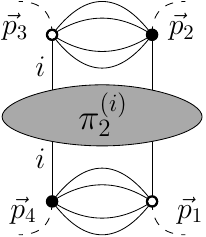} }}\,,\label{decomp4}
\end{equation}
Where the half dotted edges correspond to the amputated external propagators, and the reduced vertex functions $\pi_2^{(i)}:\mathbb{Z}^2\to \mathbb{R}$ denotes the sum of the interiors of the graphs, excluding the external nodes and the colored edges hooked to them. In the same way, one expect that the melonic effective vertex $\Gamma_{\text{melo\,}\vec{p}_1,\vec{p}_2,\vec{p}_3,\vec{p}_4,\vec{p}_5,\vec{p}_6}^{(6),i}$ is completely determined by a reduced effective vertex $\pi_3^{(i)}:\mathbb{Z}^3\to\mathbb{R}$ hooked to a boundary configuration such as:
\begin{equation}
\Gamma_{\vec{p}_1,\vec{p}_2,\vec{p}_3,\vec{p}_4,\vec{p}_5,\vec{p}_6}^{(6),i}=\vcenter{\hbox{\includegraphics[scale=0.4]{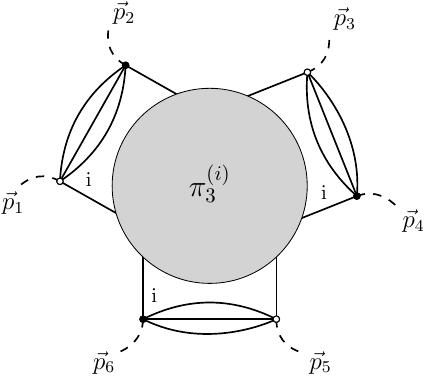} }}\,\,\, +\,\perm\,,
\end{equation}
and so one for $\Gamma^{(n),i}_{k,\,\vec{p}_1,\cdots,\vec{p}_n}$, involving the reduced vertex $\pi_n^{(i)}:\mathbb{Z}^n\to\mathbb{R}$. In the last expression, $\perm$ denote the permutation of the external edges like in \eqref{decomp4}. The reduced vertices $\pi_2^{(i)}$ can be formally resumed as a geometric series \cite{Lahoche:2018hou}-\cite{Lahoche:2018oeo}:
\begin{align}
\nonumber \pi_{2,pp}^{(1)}&= 2\left(g-2g^2 \mathcal{A}_{2,p}+4g^3(\mathcal{A}_{2,p})^2-\cdots\right)\\
&=\frac{2g}{1+2g\mathcal{A}_{2,p}}\,,\label{eff4}
\end{align}
where $ \pi_{2,pp}^{(1)}$ is the diagonal element of the matrix $ \pi_{2}^{(1)}$ and :
\begin{equation}
\mathcal{A}_{n,p}:=\sum_{\vec{p}} \,G_k^n(\vec{p}\,)\delta_{p\,p_1}\,.\label{sumA}
\end{equation}
The reduced vertex $\pi_{2,pp}^{(1)}$ depend implicitly on $k$, and the renormalization conditions defining the \textit{renormalized coupling} $g_r$ are such that:
\begin{equation}
\pi_{2,00}^{(i)}\vert_{k=0}=:2g_r\,. \label{rencond}
\end{equation}
For arbitrary $k$, the zero momentum value of the reduced vertex define the effective coupling for the local quartic melonic interaction: $\pi_{2,00}^{(i)}=:2g(k)$. The explicit expression for $\pi_3^{(1)}$ may be investigated from the proposition\ref{propmelons}. The constraint over the boundaries and the recursive definition of melonic diagram enforce the internal structure pictured on Figure \ref{fig6point} below \cite{Lahoche:2019vzy}-\cite{Lahoche:2018vun}\footnote{Note that the result is differs of a factor two with respect to the results given in the reference \cite{Lahoche:2018vun}.}. Explicitly:
\begin{equation}
\pi_{3,ppp}^{(i)}=2(\pi_{2,pp}^{(i)})^3\,\mathcal{A}_{3,p}\,, \label{6pp}
\end{equation}
The two orientations of the external effective vertices being took into account in the definition of $\pi_{2,pp}^{(i)}$.
\begin{figure}
\includegraphics[scale=0.4]{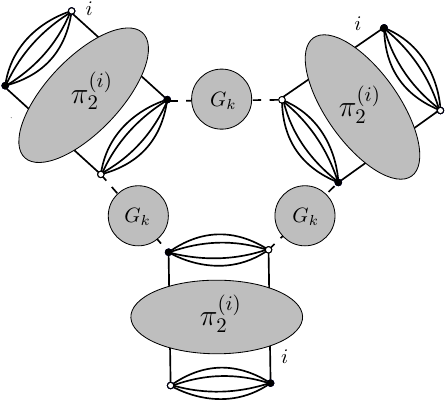}
\caption{Interne structure of the 1PI $6$-points graphs. } \label{fig6point}
\end{figure}
Expanding the exact flow equation \eqref{Wett}, and keeping only the relevant contraction for large $k$, one get the following relevant contributions for $\dot\Gamma_{k}^{(2)}$ and $\dot\Gamma_{k}^{(4)}$ :
\begin{equation}
\dot\Gamma_{k}^{(2)}(\vec{p})= -\,\sum_{i=1}^d\,\vcenter{\hbox{\includegraphics[scale=0.65]{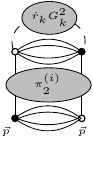} }}\label{florence}
\end{equation}
\begin{equation}
\dot\Gamma_{k}^{(4),i}=-\,\,\vcenter{\hbox{\includegraphics[scale=0.45]{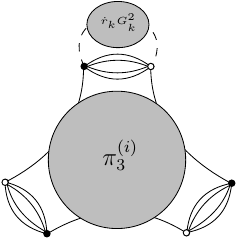} }}+\,\, 4\,\vcenter{\hbox{\includegraphics[scale=0.45]{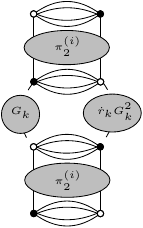} }}\label{florence2}
\end{equation}
where $\dot X:=k \partial X/\partial k$. The computation require the explicit expression of $\Gamma^{(2)}_k$. In the melonic sector, the self energy obey to a closed equation, reputed difficult to solve. We approximate the exact solution by considering only the first term in the derivative expansion in the interior of the windows of momenta allowed by $\dot r_k$:
\begin{equation}
\Gamma_{k}^{(2)}(\vec{p}\,):=Z(k)\vec{p}\,^2+m^2(k)\,,\label{derivexp}
\end{equation}
where $Z(k):=\partial\Gamma_{k}^{(2)}/\partial p_1^2(\vec{0}\,)$ and $m^2(k):=\Gamma_{k}^{(2)}(\vec{0}\,)$ are both renormalized and effective field strength and mass. From the definition \eqref{regulator}, and with some calculation (see \cite{Lahoche:2018oeo}), we obtain the following statement\footnote{Note that these equations differ from those obtained in the references by a factor $6$ in front of the sixtic contribution, for the beta function $\beta_g$.}
\begin{proposition}
In the UV domain $\Lambda\ll k \ll 1$ and in the symmetric phase, the leading order flow equations for essential and marginal local couplings are given by:
\begin{align}
\left\{
\begin{array}{ll}
\beta_m&=-(2+\eta)\bar{m}^{2}-10 \bar{g}\,\frac{\pi^2}{(1+\bar{m}^{2})^2}\,\left(1+\frac{\eta}{6}\right)\,, \\
\beta_{g}&=-2\eta \bar{g}+4\bar{g}^2 \,\frac{\pi^2}{(1+\bar{m}^{2})^3}\,\left(1+\frac{\eta}{6}\right)\Big[1\\
&\quad-6\pi^2\bar{g}\left(\frac{1}{(1+\bar{m}^{2})^2}+\left(1+\frac{1}{1+\bar{m}^{2}}\right)\right)\Big]\,. \label{syst3}
\end{array}
\right.
\end{align}
With:
\begin{equation}
\eta=4\bar{g}\pi^2\frac{(1+\bar{m}^{2})^2-\bar{g}\pi^2(2+\bar{m}^{2})}{(1+\bar{m}^{2})^2\Omega(\bar{g},\bar{m}^{2})+2\frac{(2+\bar{m}^{2})}{3}\bar{g}^2\pi^4}\,,\label{eta1}
\end{equation}
and
\begin{equation}
\Omega(\bar m^2,\bar g):=(\bar m^2+1)^2-\pi^2\bar g\,.
\end{equation}
\end{proposition}
Where in this proposition $\beta_g:=\dot{\bar g}$, $\beta_m:=\dot{\bar{m}}^2$ and the effective-renormalized mass and couplings are defined as: $\bar{g}:=Z^{-2}(k)g(k)$ and $\bar{m}^2:=Z^{-1}(k)k^{-2}m^2(k)$. For the computation, note that we made use of the approximation \eqref{derivexp} only for absolutely convergent quantities, and into the windows of momenta allowed by $\dot{r}_k$. As pointed out in \cite{Lahoche:2018hou}-\cite{Lahoche:2018oeo}, taking into account the full momentum dependence of the effective vertex $\pi_2^{(i)}$ in \eqref{eff4} drastically modify the expression of the anomalous dimension $\eta$ with respect to crude truncation. In particular, the singularity line discussed in \cite{Lahoche:2018ggd} disappears below the singularity $\bar{m}^2=-1$. Moreover, because all the effective melonic vertices only depend on $\bar{m}^2$ and $\bar{g}$, any fixed point for the system \eqref{syst3} is a global fixed point for the melonic sector. Note that to compute $\eta$, we required the knowledge of the derivative of the effective vertex with respect to the external momenta:
\begin{equation}
\frac{d}{dp^2}\pi_{2,pp}^{(i)}\bigg\vert_{p=0}=4g^2(k) \frac{d}{dp^2}\mathcal{A}_{2,p}\bigg\vert_{p=0}\,. \label{vertexderiv}
\end{equation}
 The last sum is a superficially convergent quantity. Following the observation of the authors of \cite{Lahoche:2018oeo}, it can be computed from \eqref{6pp} and Ward identity \eqref{W2}, or directly from  the truncation \eqref{derivexp}.
The system\footnote{Note that we missed a factor $2$ in the computation of \cite{Lahoche:2018vun,Lahoche:2018oeo,Lahoche:2018ggd}.
\begin{equation}
 \frac{d}{dp^2}\mathcal{A}_{2,p}\bigg\vert_{p=0}=-\frac{1}{Z^2k^2}\frac{\pi^2}{1+\bar{m}^2}\left(1+\frac{1}{1+\bar{m}^2} \right)\,.\
\end{equation}} \eqref{syst3} admits two real fixed points solutions for $p_1:=(\bar{g}_*;\bar{m}^2_*)\approx(0.34;0.18)$ and $p_2\approx (-0.02,0.53)$. These fixed points however are expected unphysical. Indeed the first one $p_1$ has anomalous dimension $\eta_1 \approx -5.8$, under the physical bound $\eta_c=-2$ (see \eqref{physicalboundeta}) and have to be discarded. For the second fixed point $\eta_2\approx -0.4$, but it has the wrong sign with respect to the stability requirement. This result contrast with the results announced in \cite{Lahoche:2018vun,Lahoche:2018oeo,Lahoche:2018ggd} (due to the missed factor 6 in the flow equations), and with the predictions of crude truncations \cite{Lahoche:2018vun}, a disagreement that can be interpreted as an artifact of the truncation approaches. Hence, we have the following statement:

\begin{claim}
Considering only the non-branching melonic sector, the effective vertex expansion allows to close the hierarchy of RG equations, and no physically relevant fixed point solutions are found.
\end{claim}\label{claim0}

\begin{figure}
\includegraphics[scale=0.3]{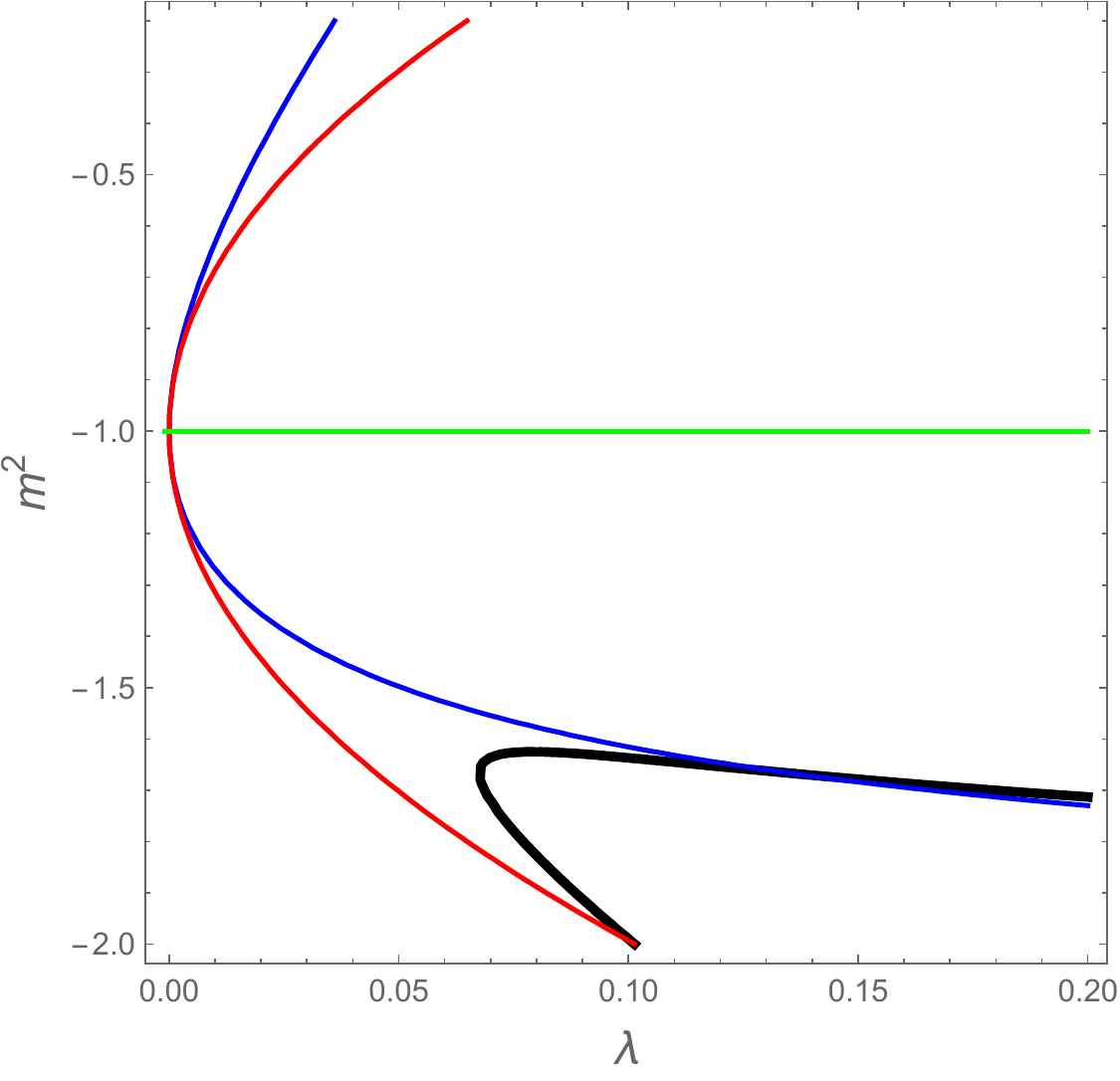}
\caption{Some relevant regions of the phase space. The curves $\eta=0$ (in blue), $(\eta)^{-1}=0$ (in black), and the curve $(\eta)^{-1}=0$ without took into account the momentum dependence of the vertex \eqref{vertexderiv} (in red). }\label{figattractor}
\end{figure}

We will see in the rest of this paper that this conclusion agrees with constraints coming from Ward identities. Note with this respect that our approximations used to compute integrals withing EVE are compatible with Ward identities, see \cite{Lahoche:2018vun,Lahoche:2018oeo,Lahoche:2018ggd} and Section \ref{COnst}. Figure \ref{figattractor} summarize the improvement coming from EVE with respect to standard vertex expansion. In particular, the singularity line (in red) of the anomalous dimension without took into account the momentum dependence of the vertex is moved to the solid black line, below the singularity $\bar{m}^2=-1$ (in green).

\section{The constrained non-branching melonic sector}\label{melons}
To close the hierarchical system derived from \eqref{Wett} and obtain the autonomous set \eqref{syst3}, we made use of the explicit expressions \eqref{eff4} and \eqref{6pp}. In this derivation, we mentioned the Ward identity but they do not contribute explicitly. In this section, we take into account their contribution and show that they introduce a strong constraint over the RG trajectories. \\

\noindent
Deriving successively the Ward identity \eqref{Ward0} with respect to external sources, and setting $J=\bar{J}=0$ at the end of the computation, we get the two following relations involving $\Gamma_k^{(4)}$ and $\Gamma_k^{(6)}$ see \cite{Lahoche:2018ggd}\footnote{Note that a factor $2$ was omitted in the first versions of the reference papers, which compensated the missed factor $2$ in front of the equation \eqref{6pp} see arXiv:1809.00247, arXiv:1809.06081, arXiv:1812.00905.} :
\begin{equation}
\pi_{2,00}^{(1)}\,\mathcal{L}_{2,k}=-\frac{\partial}{\partial p_1^2}\left(\Gamma_k^{(2)}(\vec{p}\,)-Z\vec{p}\,^2\right)\big\vert_{\vec{p}=\vec{0}}\,, \label{W1}
\end{equation}
\begin{equation}
\left(\pi_{3,00}^{(1)}\,\mathcal{L}_{2,k}-2(\pi_{2,00}^{(1)})^2\,\mathcal{L}_{3,k}\right)=-\frac{d}{dp_1^2}\pi_{2,p_1p_1}^{(1)}\big\vert_{p_1=0}\,,\label{W2}
\end{equation}
where:
\begin{equation}
\mathcal{L}_{n,k}:= \sum_{\vec{p}_\bot}\left(Z+\frac{\partial r_k}{\partial p_1^2} (\vec{p}_\bot)\right)G^n_k(\vec{p}_\bot)\,. \label{Lsum}
\end{equation}
Note that, in order to investigate the local potential approximation, we kept only the contributions with connected boundaries. We will return to this point at the end of the section \ref{COnst}. \\

It can be easily checked that the structure equations \eqref{eff4} and \eqref{6pp} satisfy the second Ward identity \eqref{W2} see \cite{Lahoche:2018hou}-\cite{Lahoche:2018oeo} and also \cite{Samary:2014tja}-\cite{Samary:2014oya}. In the same way the first Ward identity \eqref{W1} has been checked to be compatible with the equation \eqref{eff4} and the melonic closed equation for the $2$-point function. However, the last condition
does not exhaust the information contained in \eqref{eff4}. Indeed, with the same level of approximation as for the computation of the flow equations \eqref{syst3}, the first Ward identity can be translated locally as a constraint between beta functions see \cite{Lahoche:2018hou}-\cite{Lahoche:2018oeo}:
\begin{equation}
\mathcal{C}:=\beta_g+\eta\bar{g}\, \frac{\Omega(\bar{g},\bar{m}^2)}{(1+\bar{m}^2)^2}-\frac{2\pi ^2\bar{g}^2}{(1+\bar{m}^2)^3}\beta_m=0\,.\label{const}
\end{equation}
This relation hold in the deep UV limit only, that, for large $k$. Inessential contributions have been discarded, which play an important role in the IR sector $k\approx 0$, where one expect that Ward identity reduces to its unregularized form, i.e. for $r_k=0$. Moreover, note that in this limit, and depending on the choice of the regulator, it may happen that additional inessential operators have to be added to the original action to recover the true Ward identities, see Section \ref{secfin} \\

\noindent
Generally, the solutions of the system \eqref{syst3} do not satisfy the constraint $\mathcal{C}=0$. This is especially the case of the fixed point solution $p_1$ discussed previously. To be compatible with Ward identities, the fixed point should have crossed the red line in Figure \ref{figattractor}. We call \textit{constrained melonic phase space} and denote as $\mathcal{E}_{\mathcal{C}}$ the subspace of the melonic theory space satisfying $\mathcal{C}=0$.  
\medskip

\begin{figure}\label{plotf}
$\underset{a}{\vcenter{\hbox{\includegraphics[scale=0.27]{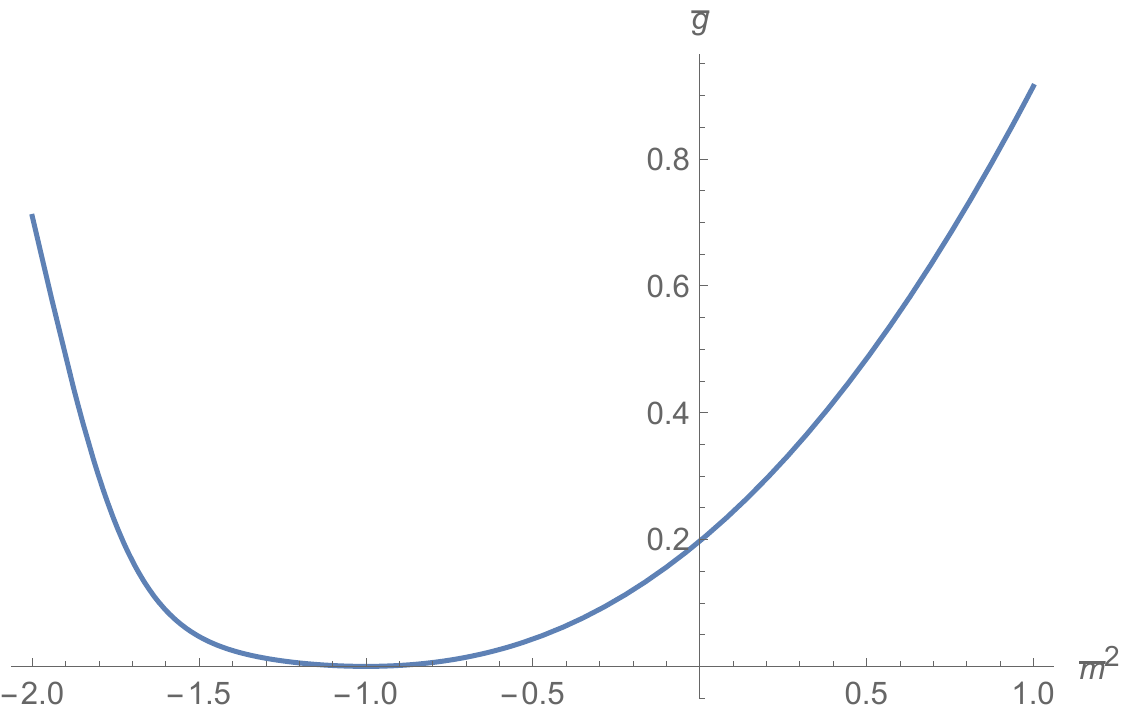} }}}$\\
$\vspace{0.5cm}$\\
$\underset{b}{\vcenter{\hbox{\includegraphics[scale=0.27]{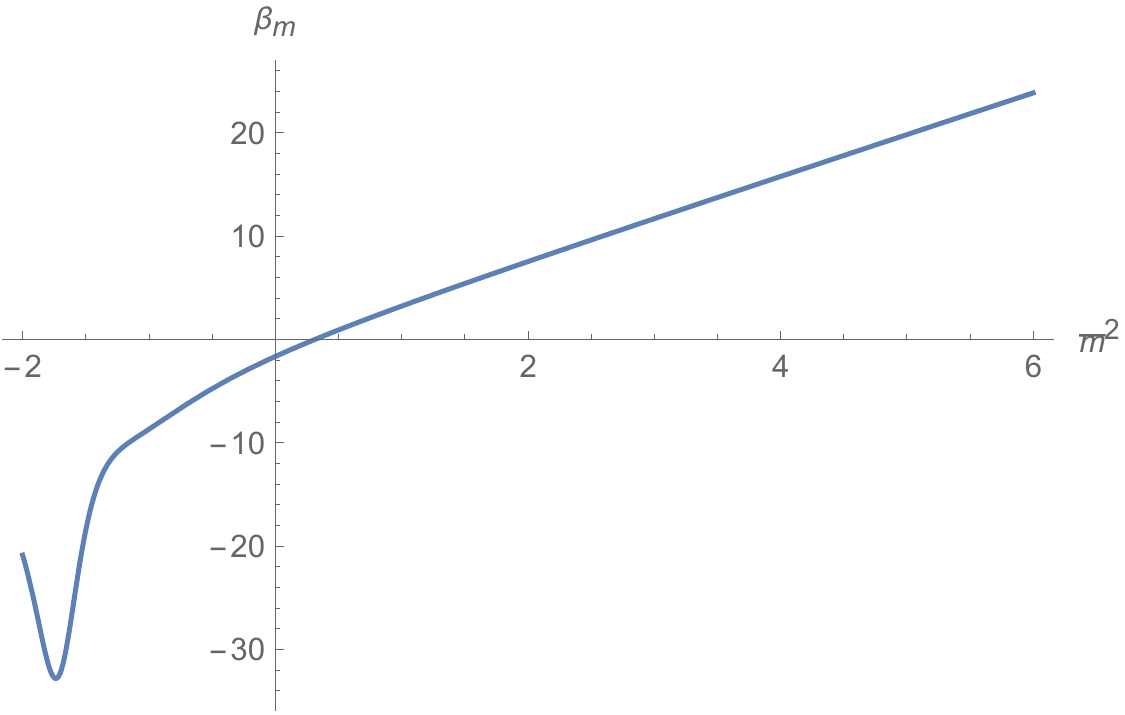} }}}$\\
$\underset{c}{\vcenter{\hbox{\includegraphics[scale=0.27]{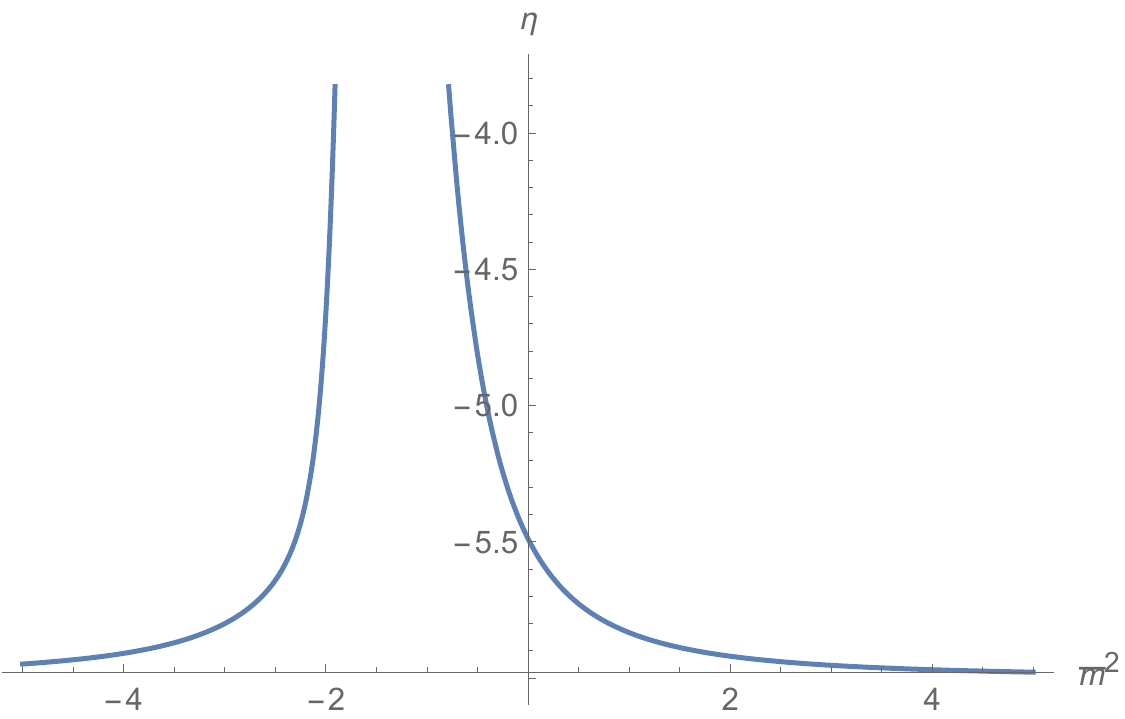} }}}$
\caption{The constrained melonic flow $\bar{g}=f(\bar{m}^2)$ ($\mathcal{E}_{\mathcal{C}1}$) in blue and $\bar{g}=0$ ($\mathcal{E}_{\mathcal{C}0}$) in brown (a); the corresponding beta function $\beta_m(\bar{m}^2,\bar{g}=f(\bar{m}^2))$ (b) and the anomalous dimension $\eta(\bar{m}^2,\bar{g}=f(\bar{m}^2))$ (c). The vertical dotted red line materializes the position of the UV attractive fixed point and the horizontal brown line materializes the physical bound $\eta=-2$.}
\end{figure}

\noindent
In the description of the physical flow over $\mathcal{E}_{\mathcal{C}}$, we substituted the explicit expressions of $\beta_g$, $\beta_m$ and $\eta$, translating the relations between velocities as a complicated constraint on the couplings $\bar{g}$ and $\bar{m}^2$ providing a systematic projection of the RG trajectories. Explicitly, solving this constraint, we get two equations for this constrained subspace $\mathcal{E}_{\mathcal{C}}$, defining respectively $\mathcal{E}_{\mathcal{C}0}$ and $\mathcal{E}_{\mathcal{C}1}$:
\begin{equation}
\bar{g}=0 \,, \quad \text{and} \quad \bar{g}=f(\bar{m}^2)\,, \label{sol}
\end{equation}
where:
\begin{equation}
f(\bar{m}^2):=\frac{(\bar{m}^2+1)^2 (15 \bar{m}^2 (\bar{m}^2+3)+37)}{\pi ^2 (\bar{m}^2 (5\bar{m}^2+19)+19)}\,,
\end{equation}
and the non-trivial solution is pictured on Figure \ref{plotf}a below. In particular $f$ has only one real zero, $\bar{m}^2=-1$, and the coupling constant in always positive. The solution $\bar{g}=f(\bar{m}^2)$ leads to the $\beta$-function and anomalous dimension:
\begin{equation}
\beta_m=\frac{4 \bar{m}^2 (\bar{m}^2 (15 \bar{m}^2 (5 \bar{m}^2 (\bar{m}^2+6)+68)+956)+163)-740}{(15 \bar{m}^2 (5\bar{m}^2 (\bar{m}^2+6)+69)+1088)\bar{m}^2+445}\,
\end{equation}
and:
\begin{equation}
\eta=-\frac{6 (5
\bar{m}^2 (\bar{m}^2+3)+11) (15 \bar{m}^2 (\bar{m}^2+3)+37)}{\bar{m}^2 (15 \bar{m}^2 (5 \bar{m}^2 (\bar{m}^2+6)+69)+1088)+445}\,.
\end{equation}
These functions are pictured on Figures \ref{plotf}b and \ref{plotf}c below. Interestingly it seems that the singularity occurring at $\bar{m}^2=-1$ in the unconstrained flow \eqref{florence} disappears, due to the factor $(1+\bar{m}^2)^2$ in the numerator of $f$. However the original flow remains undefined at this point, even if it seems allowed to continue it analytically. $\beta_m$ exhibits a fixed point solution for $\bar{m}_1^2\approx 0.32$. This fixed point is reminiscent of the fixed point $p_1$ discovered previously. Indeed, $\eta_*\approx -5.6$, and the fixed point solution is unphysical.

\section{Melonic strictly local potential approximation} \label{COnst}
Some objections can be addressed to the previous method. First of all, our computation concerns only the melonic sector, and a more complete analysis should be done, using the method explained in \cite{Lahoche:2018oeo}-\cite{Lahoche:2018vun}. Second, the computation is based on the approximation \eqref{derivexp}, keeping only the relevant terms in the derivative expansion. This approximation is used to compute the sums involved in the flow equations, into the windows of momenta allowed by $\dot{r}_k$. Note that, because this window is the same as $\partial r_k/\partial p_1^2$, no additional approximation is used to compute the Ward constraint \eqref{const}. Indeed, the undefined term:
\begin{equation}
2g \dot{\mathcal{A}}_{2,0} \,,
\end{equation}
is fixed in term of the rate $\dot{\lambda}$, from equation \eqref{eff4}:
\begin{equation}
\dot{g}=-2g^2 \dot{\mathcal{A}}_{2,0}\,.
\end{equation}
However, the computation of the flow equations has required to compute superficially convergent sums using the approximation \eqref{derivexp}, out of the windows of momenta allowed by $\dot{r}_k$. This was concerned $\pi_3$ defined by \eqref{6pp}, and the derivative of the $4$-point vertex \eqref{vertexderiv}. We checked in \cite{Lahoche:2018oeo}-\cite{Lahoche:2018vun} the compatibility of this approximation with the second Ward identity \eqref{W2}. This is a strong constraint, in favor of the reliability of our approximation. But not a good justification, without exact computation. Such an exact computation is expected to be very hard. However, we may use the constraint, not to find where our approximation makes sense in the investigated region of the full phase space, but to fix these undefined sums. This is the strategy that we will describe now, keeping the approximation \eqref{derivexp} only for the sums in the domain $\vec{p}\,^2<k^2$. As we will see, this alternative description of the interacting sector of the theory strongly simplifies the description of the constrained space, that we call $\mathcal{E}_C$ as well, and can be easily extended for models with higher-order interactions. \\

\noindent
Another objection could concern the Ward identities themselves, or more precisely the form of Ward identities that we used to compute the constraint \eqref{const}. As discussed in the previous paragraph, we discarded the effective vertices with disconnected boundaries from our analysis. However, disconnected interactions does not modify the Ward identities for connected interactions, equations \eqref{W1} and \eqref{W2}. They have their own Ward identities, but what we called Ward constraint, equation \eqref{const} holds even if we consider disconnected pieces. A simple way to check this point is to remark that, as pointed out in \cite{Lahoche:2018oeo}-\cite{Lahoche:2018vun}, the relation \eqref{W1} may be deduced from the structure of the melonic diagrams rather than a consequence of the symmetry breaking of the kinetic action. Indeed, in the melonic sector, the self energy satisfies a closed equation (see \cite{Samary:2014tja} and equation \eqref{closed}); which combined with the melonic expansion \eqref{eff4} provides exactly the same relation than \eqref{W1} (lemma \ref{lemma1} and \ref{lemma2} are specific realization of this general feature). Moreover, as explained in the previous paragraph, the second relation \eqref{W2} work as well, at least in the first order in the derivative expansion to compute the derivative of the effective vertex \eqref{vertexderiv}. Indeed, as checked in \cite{Lahoche:2018oeo}-\cite{Lahoche:2018vun}, Appendix B\footnote{Due to the missed factors $2$ recalled in footnote 3 and 5, we reproduce the proof.}:
\begin{align*}
\nonumber \frac{d}{dp^2_1}&\pi_{2,p_1p_1}^{(1)}\big\vert_{p_1=0}=-4g^2(k) \frac{d}{dp^2_1}\mathcal{A}_{2,p_1}\big\vert_{p_1=0}\\
&=8g^2(k) \sum_{\vec{p}_\bot} \left(Z(k)+\frac{\partial r_k}{\partial p_1^2}(\vec{p}_\bot)\right)G^3_k(\vec{p}_\bot) \\
&=2(\pi^{(1)}_{00})^2 \left[ \sum_{\vec{p}_\bot} \left(Z(k)-Z\right)G^3_k(\vec{p}_\bot)+\mathcal{L}_{3,k}\right]
\end{align*}
where we used \eqref{derivexp} for the second line and \eqref{Lsum} for the last line. Finally, from the first Ward identity \eqref{W1}, or mixing \eqref{eff4} and the closed equation \eqref{closed}, $(Z(k)-Z)=-Z\pi_{00}^{(1)}\mathcal{L}_{2,k}$; such that because \eqref{6pp} we recover exactly the Ward identity \eqref{W2}. Alternatively, we may view this accordance as an indication that the symmetric phase coincide with the region where the perturbative expansion converge. \\

The reliability of our approximations for the Ward constraint, however, cannot be extended for the RG flow equations themselves, and in the next two subsections, we investigate two heuristic ways to discuss the robustness of our conclusions.

\subsection{Influence of the number of interactions}

A simple way of investigation for the robustness of our conclusions about the constrained phase space is to reduce the number of interactions. Indeed, configurations as pictured on Figure \eqref{figboundary} disappears if we restrict the number of interactions to one, replacing the original model \eqref{int} by:
\begin{equation}
S_{\text{int}}[T,\bar T]= g \vcenter{\hbox{\includegraphics[scale=0.8]{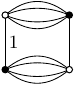} }}\,.\label{int2}
\end{equation}
In the point of view of the EVE method, equations \eqref{syst3}, the only change with respect to the fully interacting model concern the beta function $\beta_m$, the factor $10$ being in fact $2\times d$, for $d=5$. For a single color, we find two real fixed point solutions again:
\begin{equation}
p_1\approx (0.27,0.04)\,,\qquad p_2\approx (-0.009,0.08)\,.
\end{equation}
The second one has the wrong sign for the quartic coupling constant and the first one having anomalous dimension smaller than the physical bound $\eta_c$. Hence no reliable fixed point is found. 
\medskip

\begin{figure}
$\underset{a}{\vcenter{\hbox{\includegraphics[scale=0.27]{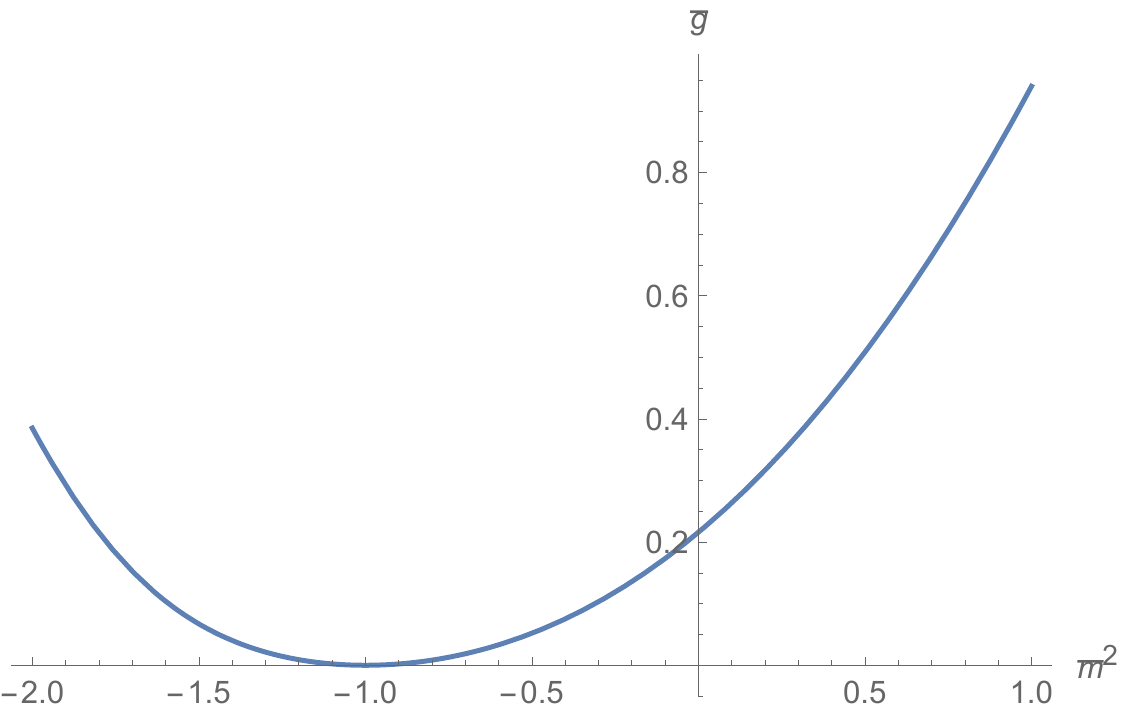} }}}$\\
$\vspace{0.5cm}$
$\underset{b}{\vcenter{\hbox{\includegraphics[scale=0.27]{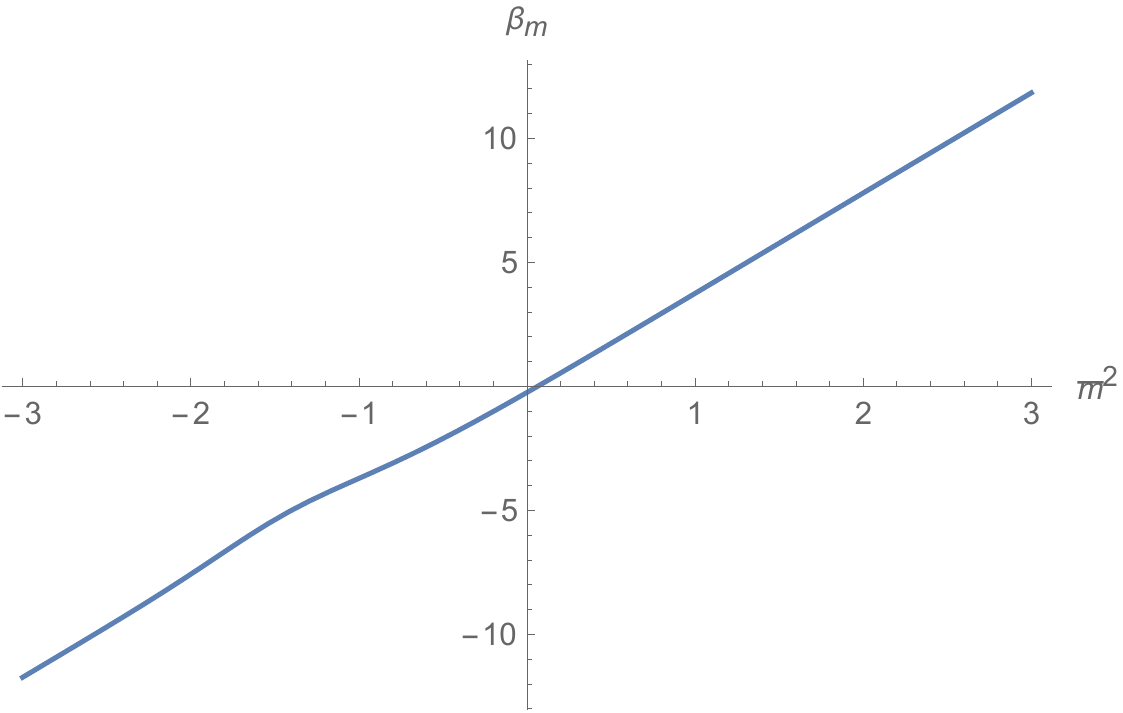} }}}$
\caption{The constrained melonic flow $\bar{g}=f(\bar{m}^2)$ ($\mathcal{E}_{\mathcal{C}1}$) in blue and $\bar{g}=0$ ($\mathcal{E}_{\mathcal{C}0}$) in brown for a single interaction (a); the corresponding beta function $\beta_m(\bar{m}^2,\bar{g}=f(\bar{m}^2))$ (b).}\label{fig1color}
\end{figure}

\begin{figure}
\includegraphics[scale=0.27]{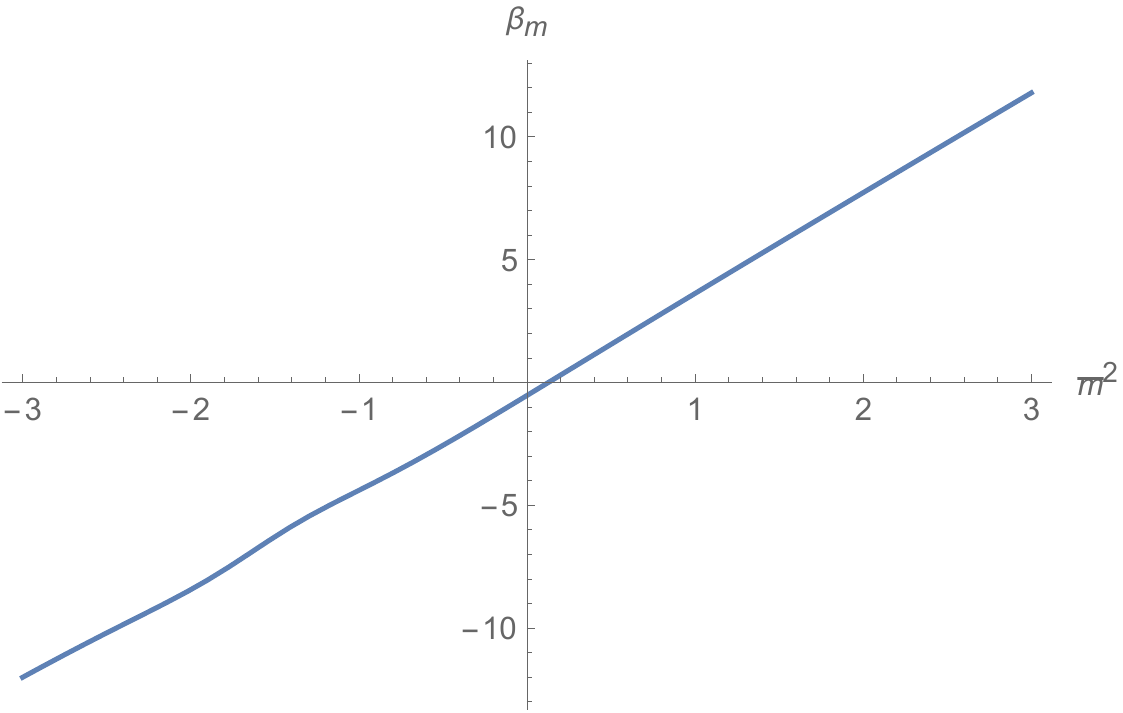}
\caption{The beta function $\beta_m$ over the constrained phase space for two interactions.}\label{betamd2}
\end{figure}

\begin{figure}
$\underset{a}{\vcenter{\hbox{\includegraphics[scale=0.27]{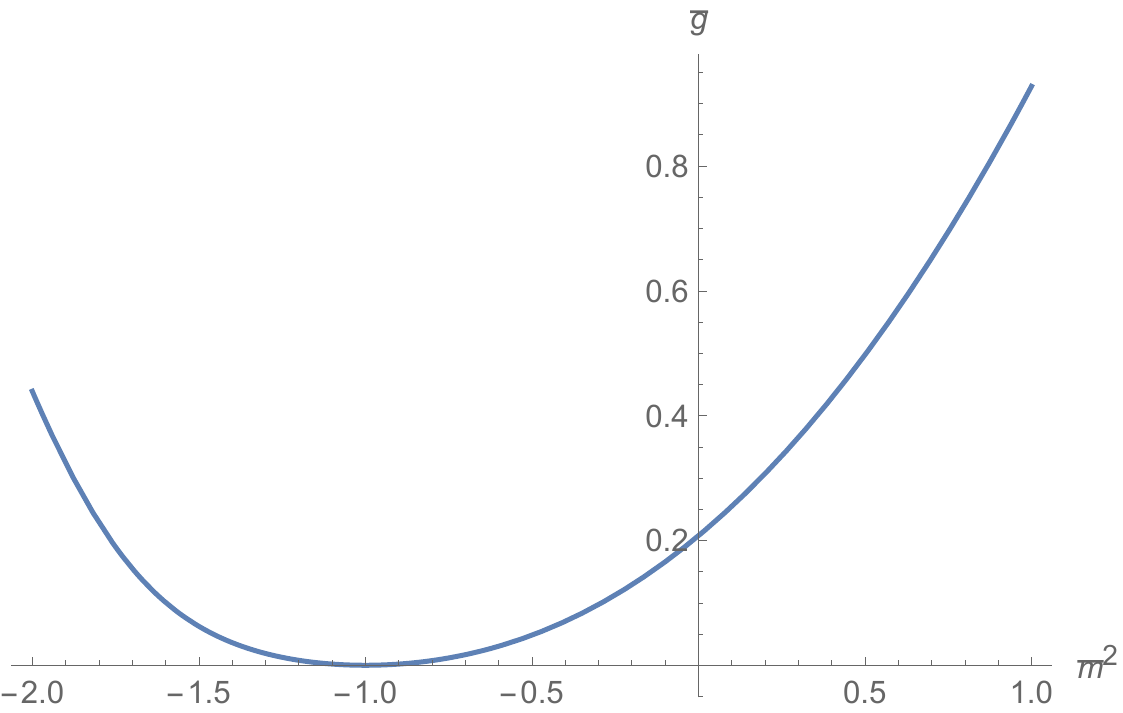} }}}$\\
$\vspace{0.5cm}$\\
$\underset{b}{\vcenter{\hbox{\includegraphics[scale=0.27]{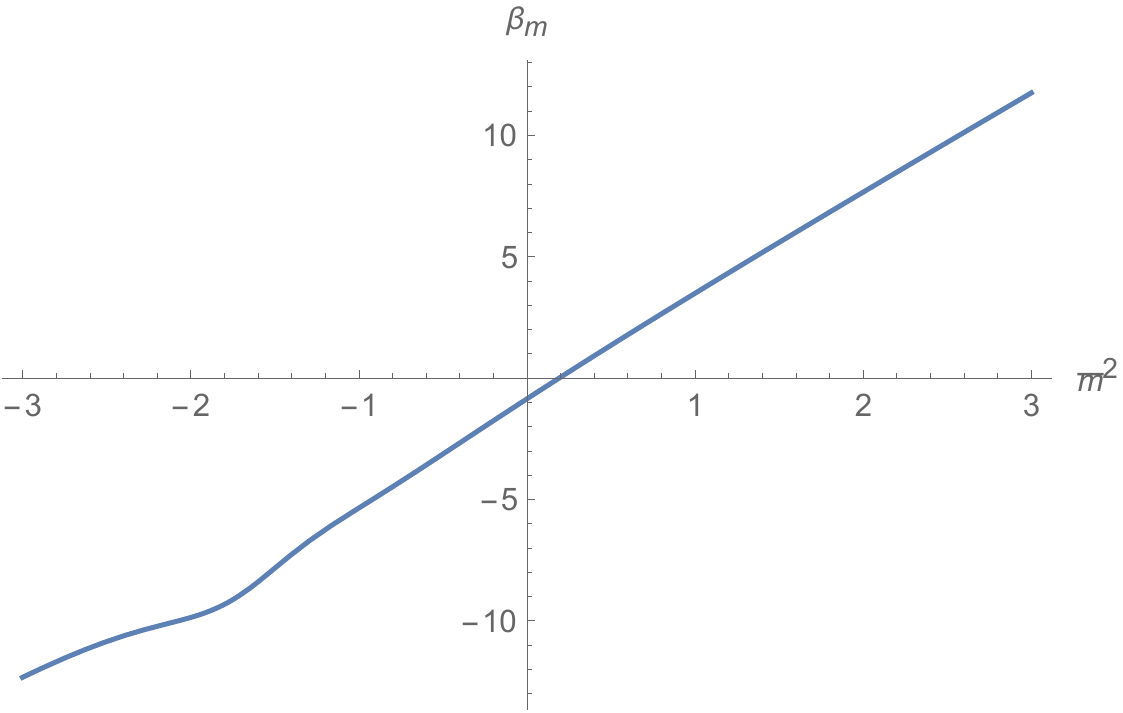} }}}$\\
$\underset{c}{\vcenter{\hbox{\includegraphics[scale=0.27]{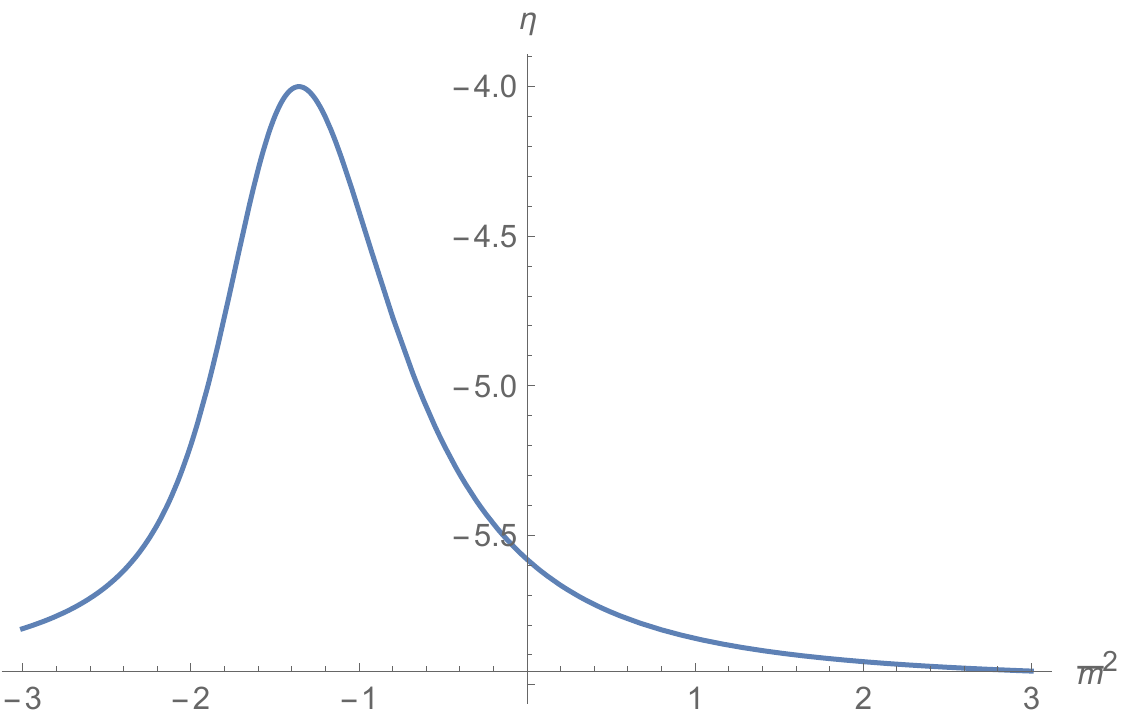} }}}$
\caption{The constrained melonic flow $\bar{g}=f(\bar{m}^2)$ ($\mathcal{E}_{\mathcal{C}1}$) in blue and $\bar{g}=0$ ($\mathcal{E}_{\mathcal{C}0}$) in brown for three interactions (a); the corresponding beta function $\beta_m(\bar{m}^2,\bar{g}=f(\bar{m}^2))$ (b); the anomalous dimension (c). }\label{figd3}
\end{figure}

The Ward constraint \eqref{const} remains unchanged, and inserting the flow equations following the strategy described in section \ref{melons}, we get the solutions pictured on Figure \ref{fig1color}. This picture have to be compared with Figure \ref{plotf}. In particular, $f$ is always positive and $\beta_m$ has the same behavior in the physical domain $\bar{m}^2>-1$ as the corresponding $\beta$-function for the fully colored model. $\beta_m$ has a single real zero, for $\bar{m}^2\approx 0.06$, which is purely attractive toward IR. It is however as unphysical as the corresponding fixed point discovered for the fully colored model, the anomalous dimension so large and negative: $\eta_*\approx -5.67$ (close to the value obtained previously). Hence, these results highlight the weak dependencies of our previous conclusions in regard to the number of quartic interactions involved to span the theory space, as soon as we focus on the isotropic sector (all interactions have the same coupling constant). In particular, one expect that disconnected interactions play no significant role. 
For $2$ interactions, the results approach the full model. $f$ remains positive and the corresponding $\beta$-function for $\bar{m}^2$ is pictured on Figure \ref{betamd2}. It has the same shape as the one-colored and the full colored model, and it has a single zero for $\bar{m}^2\approx 0.13$, which is unphysical again ($\eta_* \approx -5.6$). For $3$ interactions finally, we find again two fixed point, one of them being unstable and the other one, reminiscent of $p_1$, has negative and large anomalous dimension $\eta_* \approx -5.8$. In the constrained space, the results are summarized on Figure \ref{figd3}: We recover the main features of the fully colored model; and no reliable fixed point. To summarize:

\begin{claim}
The relevant features of the RG in the non-branching melonic sector weakly rely on the number of interactions. In particular, the predictions of the single and fully colored models are essentially the same. For this reason, we expect that disconnected interactions play no significant role in a first approximation. 
\end{claim}


\subsection{Dynamical constrained flow}

As an alternative way of investigation, we propose to fix $\pi_3^{(i)}$, and then the sum $\mathcal{A}_{3,0}$ by the flow itself, rather through a specific approximation for $\Gamma_k^{(2)}$, out of the windows of momenta allowed by $\dot{r}_k$. Our procedure is schematically the following: \\

\noindent
(1) We keep $\beta_m$ and fix $\beta_g$ from the equation \eqref{const}:
\begin{equation}
\left\{
\begin{array}{ll}
\beta_m&=-(2+\eta)\bar{m}^{2}-\,\frac{10\pi^2\bar{g}}{(1+\bar{m}^{2})^2}\,\left(1+\frac{\eta}{6}\right)\,,\\
\beta_g&=-\eta\bar{g}\, \frac{\Omega(\bar{g},\bar{m}^2)}{(1+\bar{m}^2)^2}+\frac{2\pi ^2\bar{g}^2}{(1+\bar{m}^2)^3}\beta_m\,.
\end{array}
\right.\label{systprime}
\end{equation}
(2) We fix $\pi_{3,000}^{(i)}$ dynamically from the flow equation \eqref{florence2}:
\begin{align}
\nonumber\beta_g=-2\eta \bar{g} &-3 \bar{\pi}_3^{(1)}\frac{\pi^2}{(1+\bar{m}^{2})^2}\left(1+\frac{\eta}{6}\right)\\
& +4\bar{g}^2 \,\frac{\pi^2}{(1+\bar{m}^{2})^3}\left(1+\frac{\eta}{6}\right)\label{pi3dyn}
\end{align}
(3) We compute $\frac{d}{dp_1^2}\pi_{2,00}^{(i)}$ from equation \eqref{W2}, and finally deduce an equation for the anomalous dimension $\eta$. The computation require the sums $\mathcal{L}_{2,k}$ and $\mathcal{L}_{3,k}$. $\mathcal{L}_{2,k}$ has a vanishing power counting, and contain the undefined sum $\mathcal{A}_{2,0}$. However, $\mathcal{L}_{2,k}$ may be expressed in term of $Z(k)$ and $g(k)$ from equation \eqref{W1}. Indeed, setting $k=0$ and fixing the renormalization condition such that $Z(k=0)=1$\footnote{This condition may be refined, see \cite{Lahoche:2018oeo}, but this point has no consequence on our discussion.}, we get that, in the continuum limit $\Lambda\to \infty$, $Z\to 0$. To summarize, in the same limit, \eqref{W1} reduces to $-2g(k)\mathcal{L}_{2,k}=Z(k)$, and from \eqref{W2}:
\begin{equation}
\frac{d}{dp_1^2}\pi_{2,00}^{(1)}=\left(Z(k)\frac{\pi_{3,00}^{(1)}}{2g(k)}+2(\pi_{2,00}^{(1)})^2\,\mathcal{L}_{3,k}\right)\,. \label{deriv2}
\end{equation}
To compute $\mathcal{L}_{3,k}$, we have to note that, because all the quantities are renormalized, only the superficial divergences survives. As a result, $\mathcal{A}_{3,p}$ does not diverge, such that in the continuum limit, $Z_{-\infty}\mathcal{A}_{3,p}$ must vanish, and we get straightforwardly:
\begin{equation}
\mathcal{L}_{3,k}=-\frac{1}{2Z^2(k)k^2}\frac{\pi^2}{(1+\bar{m}^2)^3}\,.
\end{equation}
Note that this term was computed using \eqref{derivexp} in the interior of the domain $\vec{p}\,^2 < k^2$. It is easy to check that the system \eqref{systprime} does not admits physically relevant fixed point using expression \eqref{eta1} for the anomalous dimension. Indeed, a fixed point have to satisfy $\eta=0$, i.e. $(1+\bar{m}^{2})^2-\bar{g}\pi^2(2+\bar{m}^{2})=0 $ or $(1+\bar{m}^{2})^2-\bar{g}\pi^2=0 $. Inserting the first condition in the flow equation for $\beta_m$, we get:
\begin{equation}
\beta_m=-2 \bar{m}^{2}-\frac{10}{2+\bar{m}^{2}}\,,
\end{equation}
which has no real zero. With the second condition however:
\begin{equation}
\beta_m=4  \bar{m}^{2}+\frac{2}{2+ \bar{m}^{2}}-6\,,
\end{equation}
which has two zeros, for $\bar{m}^2\approx -1.85$ and $\bar{m}^2\approx1.35$. The first one has to be discarded because it is located under the singularity $\bar{m}^2\approx -1$. The second one because it strongly violates the physical bound $\eta >-2$ (explicitly $\eta \approx -4.2$).
\medskip

From equation \eqref{systprime} and \eqref{pi3dyn}, we get (we omit the indices $0$ to simplify the notations) :
\begin{align*}
-&\frac{3}{\bar{g}}\pi_3^{(1)}\frac{\pi^2}{(1+\bar{m}^{2})^2}\left(1+\frac{\eta}{6}\right)=\eta+\eta\frac{\pi^2 \bar{g}}{(1+\bar{m}^2)^2}\\
&\qquad \quad -\frac{2\pi^2\bar{g}\bar{m}^2}{(1+\bar{m}^2)^3}(2+\eta)-\frac{20\pi^4\bar{g}^2}{(1+\bar{m}^2)^5}\left(1+\frac{\eta}{6}\right)\\
&\qquad \qquad -4\bar{g}\frac{\pi^2}{(1+\bar{m}^2)^3}\left(1+\frac{\eta}{6}\right)\,.
\end{align*}
Interestingly, it is easy to see, using the perturbative expansion that this formula is in accordance with the formula \eqref{6pp}. Indeed, from \eqref{eta1},
\begin{equation}
\eta\approx 4\pi^2 \bar{g} (1-2\bar{m}^2)+\cdots\,,
\end{equation}
canceling all the $\bar{g}\bar{m}^2$ terms at the leading order. Then, from equations \eqref{deriv2}, \eqref{pi3dyn} and from the flow equation \eqref{florence}, it is easy to get an explicit relation, fixing $\eta$ for a given value of the pair $(\bar{m}^2,\bar{g})$ along the flow.  After some algebraic manipulations this relation takes the form:
\begin{equation}
\eta=\frac{4 \pi ^2\bar{g}  \left(\frac{\pi ^2 \bar{g} }{5 (1+\bar{m}^2)^3}+1\right)}{(1+\bar{m}^2)^2-\Omega_1(\bar{m}^2,\bar{g})}\,, \label{stateeta}
\end{equation}
where:
\begin{equation}
\Omega_1(\bar{m}^2,\bar{g}) :=\frac{6 \pi ^2\bar{g} }{5}-\frac{4 \pi ^4 \bar{g}^2}{(1+\bar{m}^2)^3}-\frac{12 \pi ^2 \bar{g}  \bar{m}^2}{5 (1+\bar{m}^2)}-\frac{4 \pi ^2 \bar{g} }{5 (1+\bar{m}^2)}\,.
\end{equation}
 Using equation \eqref{stateeta} for $\eta$, we find that no global real fixed point survives for the system \eqref{systprime}. Hence, fixing $\pi_3$ in this way, we may expect to improve the computations of the sums implied in EVE, keeping contributions which have to be treated on the same footing with respect to the power counting.
\medskip

 Alternatively, we may assume validity of $\beta_\lambda$ and $\eta$ given by system \eqref{syst3} and \eqref{eta1}. As we pointed out these equations are compatible with Ward identities \eqref{W2}, and we may assume that missing disconnected interactions, which are essentially relevant for mass is a more serious source of physical disagreement that to use derivative expansion to compute the sums involved in the EVE. With this respect, we should define $\beta_m$ from equation \eqref{const} rather than from the flow equation. The resulting system satisfy Ward identities, up to less relevant effects than $\phi^6$ melonic interaction in regard to the power counting, and  no physically reliable fixed point solutions are found, in agreement with the previous results \ref{claim0}
\medskip

Equation \eqref{stateeta} determine $\eta$ in the constrained space $\mathcal{E}_{\mathcal{C}}$ excepts for $\bar{g}=0$ and  $\Omega_1=(1+\bar{m}^2)^2$, where it is undefined. In the vicinity of the Gaussian fixed point, we recover the universal one-loop asymptotic freedom:
\begin{equation}
\eta\approx 4 \pi^2\bar{g}\,, \qquad \beta_g=-4 \pi^2\bar{g}^2\,. \label{div}
\end{equation}
All these results are in agreement with the claim \ref{claim0}: \textit{No physically relevant fixed point exist for the model that we consider in the non-branching sector.}

\section{Another point of view on the constraint} \label{secfin}

\noindent
The existence of such strong constraints in the theory space can seems to be very surprising. Indeed, it is natural to expect that initial conditions can be chosen freely in the theory space, but our result seems to indicate that is wrong. To be more precise, we have to keep in mind that $\bar{g}(k)$ and $\bar{m}^2(k)$ are the renormalized couplings at scale $k$, as it is clear, for instance, from the renormalization condition \eqref{rencond} in the deep infrared limit. In the deep UV limit, that is to say, for $k\to \Lambda$, where the initial conditions are chosen; $\bar{g}(\Lambda)$ and $\bar{m}^2(\Lambda)$ reduce to the bare couplings $g_b$ and $m_b^2$, including non-trivial counter-terms to ensures UV finiteness of the theory. Indeed, the theory being just-renormalizable, and their exist a set of three counter-terms, $Z_m$, $Z_0$ and $Z_g$ such that for the quantum model defined with the classical action \footnote{In this section we left the regulator $r_k$.} :
\begin{align*}
S_{\text{r}}=\sum_{\vec{p}\in\mathbb{Z}^d} \bar{T}_{\vec{p}}&\left(Z_0\vec{p}\,^2+Z_{m}m^2_r\right) {T}_{\vec{p}} +Z_gg_r \sum_{i=1}^d \vcenter{\hbox{\includegraphics[scale=0.8]{phi4.pdf} }}\,,
\end{align*}
all the UV divergences, in the $\Lambda\to \infty$ limit have to be removed; the subscript ‘‘r" being for ‘‘renormalized". The initial conditions are therefore $\bar{m}^2(\Lambda)=Z_{m}m^2_r$ and $\bar{g}(\Lambda)=Z_gg_b$, such that the constraint on the initial condition simply reveal the existence of non-trivial relations between counter-terms, holding to all order in the perturbative expansion. Such a relation has been investigated in
\cite{Lahoche:2019vzy}, \cite{Lahoche:2018ggd}-\cite{Lahoche:2018oeo}, and to make this section clear, we begin to repeat shortly the arguments of the reference.\\

In the melonic sector, the $2$-point self energy $\Sigma(\vec{p}\,)$ decomposes as a sum over colors:
\begin{equation}
\Sigma(\vec{p}\,)=: \sum_{i=1}^d \tau(p_i)\,,
\end{equation}
where $\tau(p)$ satisfy a complicated closed equation:
\begin{equation}
\tau(p)=-2Z_g g_r \sum_{\vec{q}}\, \delta_{q_1p}\, \frac{1}{Z_0 \vec{q}\,^2+Z_m m_b^2-\sum_i \tau(q_i)}\,. \label{closed}
\end{equation}
Even if since this point we explicitly mention the existence of an UV cutoff $\Lambda$, referring to the deep UV sector, we follows \cite{Lahoche:2018vun}, and regularize our sums using dimensional regularization rather than a crude cutoff in momentum space, which introduce non-trivial boundary contributions. Basically, the strategy is to change $U(1)$ for $[U(1)]^D$, and to use the analyticity of the melonic amplitude with respect to the group-dimension $D$. \\

\noindent
Deriving the closed equation with respect to $p_1^2$, and setting $p_1=0$, we get:
\begin{equation}
\tau^\prime= 2Z_g g_r \mathcal{A}_{2,0}(Z_0-\tau^\prime)\,,\label{deriv1}
\end{equation}
where $\tau^\prime \equiv \tau^\prime(0)$, and where we used of the definition \eqref{sumA} for $\mathcal{A}_{2,0}$. Explicitly, the $2$ point function for small momenta writes as:
\begin{equation}
\Gamma^{(2)}_k(\vec{p}\,)=(Z_0-\tau^\prime) \vec{p}\,^2+(m^2_b-d\tau)+\mathcal{O}(\vec{p}\,^2)\,,
\end{equation}
where $\tau\equiv \tau(0)$. Then, if we fix the renormalization conditions such that, for small $\vec{p}\,^2$, $\Gamma^{(2)}_k(\vec{p}\,)\sim \vec{p}\,^2+m_r^2$, we get the relations between bare and renormalized quantities:
\begin{equation}
m_b^2:=m^2_r+d\tau\,,\qquad Z_0=1+\tau^\prime\,,
\end{equation}
such that, from \eqref{deriv1}
\begin{equation}
Z_0=1+ 2Z_g g_r \mathcal{A}_{2,0}\,. \label{Z0}
\end{equation}
The $4$-point melonic graphs may be expressed in term of the $2$-point function in the same way, using the recursive definition of melonic diagrams. Using the notation of the section \ref{EVE}, like for equation \eqref{eff4}, we get:
\begin{equation}
\pi^{(1)}_{00}=\frac{2 Z_g g_r}{1+2Z_g g_r \mathcal{A}_{2,0}}\equiv 2g_r\,,
\end{equation}
the last equality being a renormalization condition, fixing $Z_g$:
\begin{equation}
Z_g= \frac{1}{1-2g_r \mathcal{A}_{2,0}}\,,
\end{equation}
and from \eqref{Z0}, it is easy to check that :
\begin{equation}
Z_g=Z_0\,. \label{sr}
\end{equation}
As a result, the wave function and the vertex counter-term has to be equals, to all orders of the perturbative expansion. This relation can be turned in a renormalization point of view, considering the way to which the finite parts of the counter-terms move, in order to keep the relation \eqref{sr} rigid, and compare them to the Ward constraint \eqref{const}.

\subsection{Minimal subtraction}

In a first time, and as an illustration, let us consider the minimal subtraction scheme, where only the divergent parts of the divergent functions are subtracted. For $2$ point functions, we denote them respectively as $\tau_\infty$ and $\tau^\prime_\infty$. In that RG scheme, mass and wave function counter-terms are then defined as :
\begin{equation}
Z_0=1+\tau^\prime_\infty\,,\qquad Z_m=1+d\tau_\infty/m_r^2\,. \label{defminimal}
\end{equation}
Let us denote as $G_r(\vec{p}\,)$ the renormalized $2$-point function, related to the bare $2$-point function as,
\begin{equation}
G_r(\vec{p}\,)=\frac{1}{Z_0 \vec{p}\,^2+Z_mm_r^2-\Sigma(\vec{p}\,)}=\frac{1}{\vec{p}\,^2+m_r^2-\sum_i\tau_r(p_i)}\,,
\end{equation}
where the function $\tau_r(\vec{p}\,)$ has no divergences in the limit $D\to 1$, $\tau_r(\vec{p}\,)-d\tau_\infty-\tau^\prime_\infty$, and for small $\vec{p}$, the effective mass and field strength renormalization, $z_{m}$ and $z_{0}$ are such that:
\begin{equation}
G_r(\vec{p}\,) \sim \frac{1}{z_0 \vec{p}\,^2+z_m m_r^2}\,,
\end{equation}
explicitly: $z_0:= 1-\tau_r^\prime$ and $z_m m_r^2:= m_r^2-d \tau_r$. In the same way, one get for $Z_g$, $Z_g^{-1}=1-2g_r \mathcal{A}_{2,\infty}$, where once again the subscript $\infty$ means that we keep only the divergent terms in the limit $D\to 1$. From that relation, and using the definitions \eqref{defminimal} and \eqref{deriv1}, we see that even in this regularization the relation $Z_g=Z_0$ hold. Moreover:
\begin{equation}
\pi_{00}^{(1)}= \frac{2g_r}{1+2g_r \mathcal{A}_{2,r} }\,,\label{pi00}
\end{equation}
with $\mathcal{A}_{2,r}:= \mathcal{A}_{2,0} - \mathcal{A}_{2,\infty}$. The effective coupling $g_{\text{eff}}$ is then defined as:
\begin{equation}
g_{\text{eff}}:= z^{-2}_0 \frac{1}{2} \pi_{00}^{(1)}\,. \label{geff}
\end{equation}
Finally, we have the following statement:
\begin{lemma}\label{lemma1}
The finite field strength $z_0$ is equal to the finite vertex renormalization:
\begin{equation}
\pi^{(2)}_{00}=2g_r z_0\,.
\end{equation}
\end{lemma}
\noindent
\textit{Proof.} Let us start from the equation for $\tau^\prime$ \eqref{deriv1}. Splitting into infinite and renormalized parts:
\begin{equation}
\tau^\prime_{\infty}+\tau^\prime_r=2Z_g g_r (\mathcal{A}_{2,\infty}+\mathcal{A}_{2,r})(1-\tau_r^\prime)\,,
\end{equation}
where we replaced $Z_0-\tau^\prime=1-\tau_r^\prime$. From the definition \eqref{defminimal} of the counter-terms, we can rewrite the equation as:
\begin{equation}
Z_0-z_0=2Z_g g_r (\mathcal{A}_{2,\infty}+\mathcal{A}_{2,r})z_0\,.
\end{equation}
Using the equality $Z_0=Z_g$, and dividing the previous equation with respect to $Z_g$, we get:
\begin{equation}
1=\left(Z_g^{-1}+2 g_r (\mathcal{A}_{2,\infty}+\mathcal{A}_{2,r})\right)z_0\,.
\end{equation}
From the explicit expression of $Z_g^{-1}$, we cancel the divergent term, to keep:
\begin{equation}
1=\left(1+2 g_r \mathcal{A}_{2,r}\right)z_0\,.
\end{equation}
Finally, from \eqref{pi00}, we proved the lemma.
\begin{flushright}
$\square$
\end{flushright}

\subsection{Renormalization group equations}

In this section we investigate the renormalization group properties of the melonic sector through Callan–Symanzik (CS) formalism \cite{Callan:1970yg}-\cite{Symanzik:1971vw}.\\

\noindent
In the regularization procedure, the coupling take a nonzero dimension, and scales like $\mu^{2(1-D)}$ for some referent scale $\mu$ used to fix the renormalization conditions; and the dependence of the effective coupling on the finite part of the counter-terms imply the existence of a global parametrization invariance:
\begin{equation}
\Gamma_\mu^{(N)}(m^2(\mu),g(\mu))\approx \frac{Z^{N/2}(\mu)}{Z^{N/2}(\mu^\prime)} \Gamma_{\mu^\prime}^{(N)}(m^2(\mu^\prime),g(\mu^\prime))\,.
\end{equation}
This global parametrization invariance may be translated as a partial differential equation, which writes as:
\begin{equation}
\left[\frac{\partial }{\partial t}+\beta(\mu) \frac{\partial}{\partial g}+\gamma_m(\mu) m^2\frac{\partial}{\partial m^2}+\frac{N}{2}\gamma(\mu)\right]\Gamma_\mu^{(N)}= 0\,. \label{RGequation}
\end{equation}
To a given subtraction scale $\mu$, the renormalization conditions are fixed such that:
\begin{equation}
\tau^\prime(p=\mu)=0\,,\qquad \pi_{\mu\mu}^{(2)}=2g(\mu)\,. \label{renscalemu}
\end{equation}
As a consequence, from the melonic structure equations, the following statement holds:
\begin{proposition}\label{propheart}
To all order of the perturbative expansion, and in the deep UV sector, the anomalous dimension $\gamma(\mu)$ and the beta function $\beta(\mu)$
\begin{equation}
\dot{g}(\mu)\equiv \beta(g(\mu))=-\gamma(\mu)g(\mu)\,. \label{betaconst}
\end{equation}
\end{proposition}

\noindent
\textit{Proof.} To prove this proposition, let us define$ \chi(\mu):=\ln(z(\mu))$. From the renormalization conditions:
\begin{equation}
\pi_{pp}^{(2)}=\frac{2g(\mu)}{1+2g(\mu)(\mathcal{A}_{2,p}-\mathcal{A}_{2,\mu})}\,,
\end{equation}
fixing the counter term $Z_g$ as :
\begin{equation}
Z_g^{-1}=(g_r/g(\mu))(1-2g(\mu) \mathcal{A}_{2,\mu})\,. \label{ZZg}
\end{equation}
To fix the mass and wave function counter-terms, let us expand the unrenormalized $2$-point function in the vicinity of $p_i=\mu$:
\begin{align}
\nonumber G_\mu^{-1}(\vec{p}\,)=(Z_0-\tau^\prime(\mu))&\vec{p}\,^2+(Z_mm_r^2-d\tau(\mu)\\
&+d\mu^2 \tau^\prime(\mu))-\sum_i \tau_\mu(p_i)\,,
\end{align}
where we defined the subtracted self energy $\tau_\mu(p)$ as
\begin{equation}
\tau_\mu(p)=\tau(p)-\tau(\mu)-(p^2-\mu^2)\tau^\prime(\mu)\,.
\end{equation}
From the renormalization conditions \eqref{renscalemu}, it follows that $Z_0-\tau^\prime(\mu)$ have to be equal to $1$, and we define the effective mass at scale $\mu$ as:
\begin{equation}
m_\mu^2:=Z_m m_r^2-d\tau(\mu)+d\mu^2 \tau^\prime(\mu)\,.
\end{equation}
It is however more convenient to fix the renormalization condition for vanishing momentum, that is, at scale $\mu=0$:
\begin{equation}
G_\mu^{-1}(\vec{p}=\vec{0}\,)=:m^2(\mu)=m_\mu^2-d\tau_\mu(0)\,,
\end{equation}
depending on $\mu$ through the other renormalized parameters, and where we used the definition of $\tau_\mu(p)$ for the last equality. Expanding $G_\mu^{-1}(\vec{p}\,)$ in the vicinity of $\vec{p}=0$, we then get:
\begin{equation}
G_\mu^{-1}(\vec{p}\,)\approx z(\mu)\vec{p}\,^2+m^2(\mu)+\mathcal{O}(\vec{p}\,^2)\,,
\end{equation}
with:
\begin{equation}
z(\mu)=1-\tau^\prime_\mu(0)\,,
\end{equation}
and we have the equivalent-lemma of the lemma \ref{lemma1}:
\begin{lemma} \label{lemma2}
For arbitrary renormalization conditions at scale $\mu$; the effective wave function $z(\mu)$ and the effective melonic vertex $\pi^{(0)}_{00}$ are related as:
\begin{equation}
\pi^{(0)}_{00}=2g(\mu)z(\mu)\,.
\end{equation}
\end{lemma}

\noindent
\textit{Proof.} The proof follows the one of the lemma \ref{lemma1}. From the equation \eqref{deriv1}, setting $p=\mu$, we get:
\begin{equation}
\tau^\prime(\mu)=2 Z_g g_r \mathcal{A}_{2,\mu}(Z_0-\tau^\prime(\mu))\,.
\end{equation}
The term in brackets on the right hand side is equal to $1$ from the renormalization condition. Therefore:
\begin{equation}
Z_0=1+2 Z_g g_r \mathcal{A}_{2,\mu} \Rightarrow Z_g=\frac{g(\mu)}{g_r} Z_0\,. \label{ZgZ0}
\end{equation}
Then, for $p=0$,
\begin{equation}
\tau_\mu^\prime(0)=\tau^\prime(0)-\tau^\prime(\mu)=1+\tau^\prime-Z_0\,,
\end{equation}
leading to:
\begin{equation}
\tau^\prime(0)=Z_0-z(\mu)=2Z_g g_r \mathcal{A}_{2,0}(Z_0-\tau^\prime(0))\,.
\end{equation}
From definition of $z(\mu)$, and because of the relation between $Z_g$ and $Z_0$ given by \eqref{ZgZ0}, we get:
\begin{equation}
1=z(\mu)\left[1/Z_0+2g(\mu) \mathcal{A}_{2,0} \right]\,.
\end{equation}
Multiplying the two hands by $2g(\mu)$, and using equations \eqref{ZZg} and \eqref{ZgZ0} to compute $1/Z_0$, we find:
\begin{equation}
2g(\mu)z(\mu)=\frac{2g(\mu)}{1+2g(\mu)(\mathcal{A}_{2,0} -\mathcal{A}_{2,\mu} )}\,.
\end{equation}
\begin{flushright}
$\square$
\end{flushright}
The basis ingredients of the proposition \eqref{propheart} are follows from this key lemma. Indeed, the CS equation \eqref{RGequation} may be concisely written as $[D+n\gamma/2]\Gamma^{(N)}_\mu=0$, $D$ denoting collectively all the linear derivatives with respect to $t$, $g$ and $m$. One one hand, for the $2$ point function, keeping only the linear terms in $\vec{p}^2$, we get $Dz(\mu)+\gamma z(\mu)=0$. One the second hand, from lemma \ref{lemma2}, we get for $\Gamma_\mu^{(4)}$:
$g(\mu)Dz(\mu)+z(\mu) Dg(\mu)+2\gamma z(\mu)g(\mu)=0$. As a result, we must has necessarily, for nonvanishing $z(\mu)$:
\begin{equation}
Dg(\mu)+\gamma g(\mu)=0\,,
\end{equation}
and $Dg(\mu)\equiv \beta(g(\mu))$.
\begin{flushright}
$\square$
\end{flushright}

\noindent
The RG coefficients $\beta$, $\gamma_m$ and $\gamma$ can be computed in terms of melonic equations. Indeed, from :
\begin{equation}
\Gamma^{(2)}(\vec{p}\,)=z_0 \vec{p}\,^2+z_m m_r^2+\mathcal{O}(\vec{p}\,^2)\,,
\end{equation}
inserted into the CS equation, we get two independent equations :
\begin{align}
\chi_{m,t} +\beta \chi_{m,g} &+\gamma_m(1+\chi_{m,m^{2}} )+\gamma =0\,,\\
\chi_{,t}+\beta \chi_{,g}&+\gamma_m\chi_{,m^{2}}+\gamma=0 \,,
\end{align}
where $\chi:= \ln(z_0)$, $\chi_m=\ln(z_m)$ and $t=\ln(\mu)$; and where we used of the notation $\partial f/\partial x=:f_{,x}$. Substituting $\beta=-\gamma g$, we can solve these equations for $\gamma$ and $\gamma_m$. What is important for our purpose is that proposition \ref{propheart} corresponds to what we called Ward constraint in the previous sections, equation \eqref{const}. The two relations are the same at one and two loops, neglecting contributions arising from mass couplings. These contributions, involving $\beta_m$ arise from the regulation procedure, as a consequence of the relevant nature of the mass coupling. Once again, in this derivation, we do not use of the Ward identity, but only of the melonic structure of the leading order graphs, highlighting the complementarity of the two approach, especially in regard to disconnected contributions, discussed at the beginning of this section.

\section{Conclusion and discussion}\label{conclusion}

In this paper, we addressed a short presentation and an extended discussion about an approximate solution of the renormalization group flow in the melonic local potential approximation. In contrast with other methods like standard vertex expansion for the most popular; effective vertex expansion allows to keep into account sector of infinite size, exploiting a parametrization of all irrelevant interactions in terms of the relevant ones. This allows in particular to close the infinite hierarchy of flow equations around the just-renormalizable sector. Investigating the fixed points solutions, we showed that no reliable, physically relevant fixed point exist in the non-branching isotropic sector, and that this conclusion weakly rely on the number of quartic interactions in the microscopic model. 
\medskip

Translating the Ward identity in the non-branching melonic sector, we proved the existence of strong relations between $\beta$-functions. These relations allow to overcome the limitations of the purely local expansion, and to avoid computation of undefined integrals as in the first version of the EVE discussed previously. Investigating the fixed point solutions for this flow agreeing with Ward identities, we confirm the disappearance of fixed point solutions, what we understand as artifact of crude truncations of the full theory space.

\onecolumngrid

\end{document}